\documentclass[aps,pra,twocolumn,superscriptaddress,notitlepage,nofootinbib,longbibliography]{revtex4-2}
\usepackage{mathrsfs}
\usepackage{epsfig}
\usepackage{graphicx}
\usepackage{amsfonts}
\usepackage{amssymb}
\usepackage{amsmath}
\usepackage{dcolumn}
\usepackage{bm}
\usepackage{xcolor}
\usepackage{braket}
\usepackage{units}
\usepackage{xspace}
\usepackage{orcidlink}

\usepackage{bbold}
\usepackage{comment}
\usepackage{soul}
\usepackage{float}
\usepackage{appendix} 

\newcommand{\ZJNU}{Department of Physics, Zhejiang Normal University, Jinhua 321004, China}
\newcommand{\JSU}{Department of Physics, Jiangsu University, Zhenjiang 212013, China}
\begin{document}
\title{Multicomponent cat states with sub-Planckian structures and their optomechanical analogues}

\author{Tan Hailin~\orcidlink{0009-0005-2531-7498}}
\affiliation{\ZJNU}
\author{Naeem Akhtar~\orcidlink{0000-0002-6756-2898}}
\email{Corresponding author: naeemakhtar166067@gmail.com}
\affiliation{\ZJNU}
\affiliation{\JSU}
\author{Gao Xianlong~\orcidlink{0000-0001-6914-3163}}
\email{gaoxl@zjnu.edu.cn}
\affiliation{\ZJNU}
\date{\today}

\begin{abstract}

We investigate the superposition of coherent states, emphasizing quantum states with distinct Wigner phase-space features relevant to quantum information applications. 
In this study, we introduce generalized versions of the compass state, which display enhanced phase-space characteristics compared with the conventional compass state, typically a superposition of four coherent states. Our findings reveal that, unlike sub-Planckian structures and phase-space sensitivity of the compass state, these generalized states produce isotropic sub-Planckian structures and sensitivity to phase-space displacements. We demonstrate that these desirable phase-space characteristics are maintained in superpositions comprising at least six distinct coherent states. Furthermore, we show that  increasing the number of coherent states in the superposition preserves these characteristics, provided the number remains even. We examine an optomechanical system capable of generating the proposed quantum states, resulting in optomechanical counterparts with nearly identical phase-space structures, thereby suggesting the feasibility of physically realizing these generalized compass states.
\end{abstract}
\maketitle

\section{Introduction}\label{sec:intro}

The concept of a macroscopic cat state~\cite{glauber_coherent_1963,dodonov_even_1974,yurke_generating_1986}, originating from a well-known gedanken experiment~\cite{schrodinger_discussion_1935}, involves the superposition of two macroscopically distinct coherent states. This idea has been extended to multicomponent cat states, which are formed by superposing three or more coherent states~\cite{janszky_coherent_1993,shukla_quantum_2019,howard_quantum_2019}. These quantum states have numerous applications in quantum information science~\cite{PRXQuantum.2.030204,ourjoumtsev_generating_2006,ourjoumtsev_generation_2007,goto_bifurcation-based_2016,sychev_enlargement_2017,cochrane_macroscopically_1999,van_enk_entangled_2001}. Compass states~\cite{zurek_sub-planck_2001,akhtar_sub-planck_2021,akhtar_sub-planck_2022,akhtar_sub-planck_2023,shukla_superposing_2023}, particularly the Zurek compass state, which is a superposition of two cat states~\cite{zurek_sub-planck_2001}, exhibit intriguing phase-space structures. The Wigner function~\cite{Sch01, CASTAGNINO2006879} of these states can reveal features at scales significantly below the Planck limit $\hbar$, known as sub-Planckian structures~\cite{Robertson1929,wheeler2014}. These sub-Planckian structures are highly susceptible to environmental decoherence~\cite{PhysRevA.109.053718,LI2002975} and play a crucial role in quantum sensing applications~\cite{toscano_sub-planck_2006,dalvit_quantum_2006}, enhancing sensitivity to phase-space displacements far below the standard limit~\cite{akhtar_sub-planck_2021,akhtar_sub-planck_2022,akhtar_sub-planck_2023}. 

The sub-Planckian phase-space structures of the Zurek compass state are inherently anisotropic~\cite{akhtar_sub-planck_2022,shukla_superposing_2023}, leading to directionally dependent sensitivity enhancements. This anisotropy causes certain directions of phase-space displacement to exhibit greater sensitivity than others. Various interpretations of compasslike states, incorporating symmetries beyond those of the harmonic oscillator algebra, have been proposed~\cite{AGUERO20002203,akhtar_sub-planck_2021,akhtar_sub-planck_2022,akhtar_sub-planck_2023}. Both theoretical and experimental techniques have been developed to create these compasslike states~\cite{agarwal_mesoscopic_2004,stobinska_wigner_2008,roy_sub-planck-scale_2009,ghosh_sub-planck-scale_2009,arman_generating_2024,vlastakis_deterministically_2013,kirchmair_observation_2013,johnson_ultrafast_2017,hastrup_deterministic_2020}.

Isotropic forms of sub-Planckian structures have been introduced~\cite{shukla_superposing_2023,akhtar2024subshotnoisesensitivitydeformed}, demonstrating superior performance over their anisotropic counterparts due to their uniform phase-space sensitivity enhancement. This isotropy offers greater potential in quantum sensing applications~\cite{akhtar2024subshotnoisesensitivitydeformed} compared with the Zurek compass state~\cite{zurek_sub-planck_2001}. While the superposition of two or more compass states results in isotropic sub-Planckian structures~\cite{shukla_superposing_2023}, we have also identified that superpositions of macroscopic cat states, including two-component cat states, can achieve similar isotropic sub-Planckian phase-space structures. Specifically, we show that the superposition of higher-order cat states produces isotropic sub-Planckian structures, aligning with existing compass-state superpositions~\cite{shukla_superposing_2023}. Additionally, we propose nearly analogous quantum states derived from an optomechanical setting, providing a pathway for the physical realization of these states.

Multicomponent cat states have been studied in various contexts~\cite{shukla_quantum_2019, KILIN199585, PhysRevA.93.062323,HOROSHKO201967,PhysRevLett.91.017902,Lee15}. In our work, we focus on a specific form of multicomponent cat states, which are obtained by superposing particular cat states to achieve the desired phase-space characteristics. Our analysis is structured as follows: We first introduce multicomponent cat states in Sec.~\ref{sec:superpositions}, followed by an analysis of their corresponding phase space using the Wigner function. We demonstrate how these states can achieve isotropic sub-Planckian structures. Specifically, we show that isotropic sub-Planckian structures generally arise from the superposition of at least three specific cat states, and this concept can be generalized to the superposition of larger numbers of specific cat states. The presence of isotropic sub-Planckian structures in cat-state superpositions indicates that these features are not exclusive to compass states~\cite{shukla_superposing_2023}. Thus, the sub-Planckian structures observed in compass-state superpositions are specific instances within the broader framework we present. Furthermore, in Sec.~\ref{subsec:sensitivity1}, we examine the phase-space sensitivity of each example, finding isotropic displacement sensitivity across all scenarios. This highlights their advantage over compass states and aligns their properties with comparable compass-state superpositions~\cite{shukla_superposing_2023}. This comparability broadens the understanding of isotropic sub-Planckness, extending the concept beyond previous investigations~\cite{shukla_superposing_2023}.

The generation of nonclassical states via cavity systems is a crucial area in quantum information science and has been extensively explored~\cite{PhysRevLett.127.087203,PhysRevA.98.063814,PhysRevApplied.21.044018,YUAN20221,LI202315,PhysRevLett.130.213604,PhysRevA.108.063505,PhysRevLett.129.037205,Zhang21,zuo2024squeezedlightexcitonphononcavity,bose_preparation_1997}. In Sec.~\ref{sec:optomechanical}, we discuss the generation of multicomponent cat states within an optomechanical framework. Notably, while this optomechanical setup was previously used to generate compass states~\cite{bose_preparation_1997}, it was not further investigated for the generation of higher-order multicomponent cat states, as achieved in our work. Our analysis is further enriched by an in-depth discussion of the impact of decoherence on the corresponding characteristics of the states.
 
 \section{Generic superposition of states}\label{sec:superpositions}
 
\begin{figure}
\includegraphics[width=0.5\textwidth]{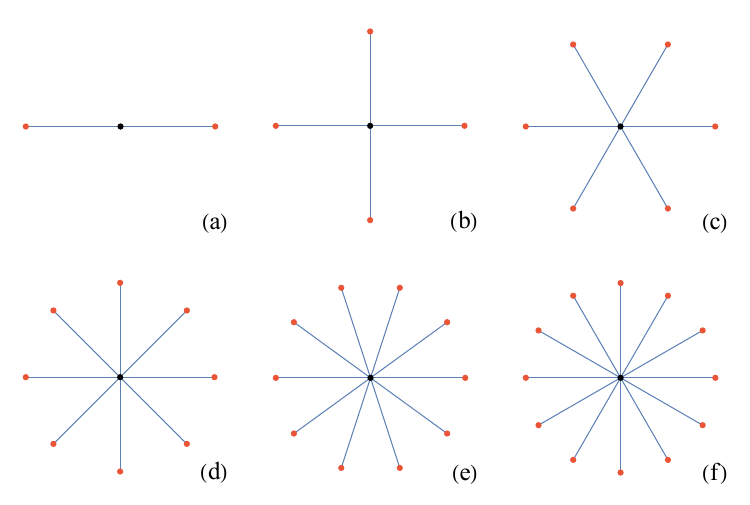}
\caption{The superposition of $L$ coherent states forms an $L$-component cat state: (a) $L=2$, (b) $L=4$, (c) $L=6$, (d) $L=8$, (e) $L=10$, and (f) $L=12$. This schematic illustrates the configurations of $L$-component cat states achievable in phase space through precise parameter selection. We focus on these specific multicomponent cat states, where red dots indicate the component coherent states in each superposition, and black dots mark a central phase-space component. Blue lines denote fixed distances from the origin, with equal spacing between neighboring coherent states, resulting in the superposition of coherent states along a circle in phase space for $L\ge 4$.}
\label{fig:fig1}
\end{figure}

The macroscopic superposition of two distinguishable coherent states leads to a macroscopic cat state~\cite{glauber_coherent_1963,dodonov_even_1974,yurke_generating_1986}. This notion is also well-known for the superposition involving more than two coherent states~\cite{zurek_sub-planck_2001, shukla_quantum_2019, shukla_superposing_2023}. Here, we examine the concept of such higher-order coherent-state superpositions. The generic superpositions of $L$ coherent states are presented in the Appendix. However, our main focus is the case when the superposition involves a number $L$ of coherent states of equal weights and particular phases~\cite{janszky_coherent_1993,toscano_sub-planck_2006,shukla_superposing_2023}:
\begin{equation}\label{eq:genric1}
\ket{\beta_L}=\frac{1}{\sqrt{\mathcal{N}_L}} \sum_{j=0}^{L-1} \ket{\mathrm{e}^{\mathrm{i} \omega_j}\beta },
\end{equation}
with $\omega_j = \nicefrac{2 \pi j}{L}$. In our work, we consider particular superpositions as illustrated in Fig.~\ref{fig:fig1}. Figure~\ref{fig:fig1}(a) depicts the superposition of two coherent states, corresponding to $L=2$ in Eq.~(\ref{eq:genric1}) with a specific phase selection. Higher-order superpositions are illustrated in Fig. ~\ref{fig:fig1}: in Fig.~\ref{fig:fig1}(b) for $L=4$, in Fig.~\ref{fig:fig1}(c) for $L=6$, in Fig.~\ref{fig:fig1}(d) for $L=8$, in Fig.~\ref{fig:fig1}(e) for $L=10$, and in Fig.~\ref{fig:fig1}(f) for $L=12$. The primary focus of the present work is on these higher-order superpositions, ranging from $L = 6$ to $L = 12$. The red-dotted patches, located away from the origin, represent each component of the corresponding coherent states, while black dots added at the origin highlight distinctive phase-space features that are the focus of our analysis. Each illustration corresponds to the specific configuration that yields the relevant $L$-component cat state in phase space. For the purposes of this work, we consider the superposition of coherent states ($L \geq 6$) that are equidistant from the origin and evenly distributed along the phase-space origin, forming a circular pattern. These particular superpositions give rise to the phase-space features of interest. Notably, in the superposition described by Eq.~(\ref{eq:genric1}), the weights of the coherent states are kept constant, and a specific choice of phases is made, representing the ideal case in this work. Any variations in the relative phases or weights could impact the phase-space features of the corresponding states, which we further explore in the Appendix.

The normalization coefficient $\mathcal{N}_L$ is denoted as
\begin{equation}\label{eq:2}
\begin{aligned}
\mathcal{N}_L=L  + \sum_{j > k}^{L-1} 2 \mathrm{e}^{-2 |\beta|^2 \sin ^2{\left(\omega_{j}-\omega_{k}\right)}}\cos{\left[|\beta|^2 \sin \left(\omega_j - \omega_k\right)\right]}.
\end{aligned}
\end{equation}
For $\beta \gg1 $, the factor $\mathrm{e}^{-2 |\beta|^2}$ in the second term in Eq.~(\ref{eq:2}) can be ignored, allowing us to approximate the normalization constant as $\mathcal{N}_{L} \simeq L$. Throughout this work, we adopt natural units by setting the reduced Planck constant $\hbar = 1$.

The Wigner function for a quantum state $\rho$ is a phase-space quasiprobability distribution, represented as~\cite{wigner_quantum_1932,groenewold_principles_1946,moyal_quantum_1949,ZAYED2005967}
\begin{equation}
W_{\hat{\rho}} \left(\alpha\right) = \frac{2}{\pi}\mathrm{Tr}[\hat{\rho}\hat{D} \left(\alpha\right)\left(-1\right)^{\hat{a}^\dagger\hat{a}}\hat{D}^\dagger \left(\alpha\right)],
\end{equation}
where $\hat{D}\left(\alpha\right)= \mathrm{e}^{\alpha \hat{a}^\dagger - \alpha^* \hat{a}}$, where $\alpha\in \mathbb{C}$ represents the displacement operator. The complex $\alpha$, with $\text{Re}(\alpha)$ and $\text{Im}(\alpha)$, defines position and momentum spaces, respectively.

To simplify our calculations, we present the generic form of the Wigner function for the operator $\ket{\beta^\prime}\bra{\beta^{\prime\prime}}$, given by
\begin{equation}\label{eq:elem_wigner}
W_{\ket{\beta^\prime}\bra{\beta^{\prime\prime}}}\left(\alpha\right) =\frac{2}{\pi} \mathrm{e}^{-\frac{1}{2}\left(\left|\beta^\prime\right|^2+\left|\beta^{\prime\prime}\right|^2\right) + \beta^\prime (\beta^{\prime\prime})^* -2\left(\alpha-\beta^\prime\right)\left(\alpha-\beta^{\prime\prime}\right)^* },
\end{equation}
which then can be used to evaluate the Wigner function of the $L$-component cat state denoted as $ W_L\left(\alpha\right)$:
\begin{equation}
W_L\left(\alpha\right) = \frac{1}{\mathcal{N}_L}\sum_{j,k=0}^{L-1} 
W_{\ket{\beta \mathrm{e}^{\mathrm{i}\omega_j}}\bra{\beta \mathrm{e}^{\mathrm{i} \omega_k}}}\left(\alpha\right).
\label{eq:L_compo_Wigner}
\end{equation}
In the scenarios illustrated in Fig.~\ref{fig:fig1}, the constituent coherent states are uniformly distributed along a circular path, each contributing equally to the phase-space extension (the phase-space area) of the central feature (dotted region). This was previously demonstrated for four-component cat states~\cite{akhtar_sub-planck_2021}, and remains valid for the higher-order superpositions examined in this work. In our cases we consider the superposition of coherent states having uniform amplitude $\beta$ throughout our analysis, and it can be easily shown that the extension of the central feature depends only on $\beta$.

The characteristics of the $L$-component cat state described in Eq.~(\ref{eq:genric1}) are discussed below, with illustrative examples shown in Fig.~\ref{fig:fig1}. This analysis explores quantum states with unique sub-Planckian structures, aiming to compare these cases to understand how their configurations affect characteristics in phase space. We focus on how each state type exhibits distinct features and behaviors, highlighting intriguing properties in each context. This comparison enhances our understanding of the fundamental principles governing these states and their interactions. Quasiclassical states, such as the coherent state of a harmonic oscillator, adhere to minimal uncertainty principles \cite{PRXQuantum.2.030204}. Sub-Planckian structures, however, occupy dimensions significantly smaller than this standard and are constrained in all directions of phase space simultaneously.

\begin{figure}[h]
\centering
\includegraphics[width=0.52\textwidth]{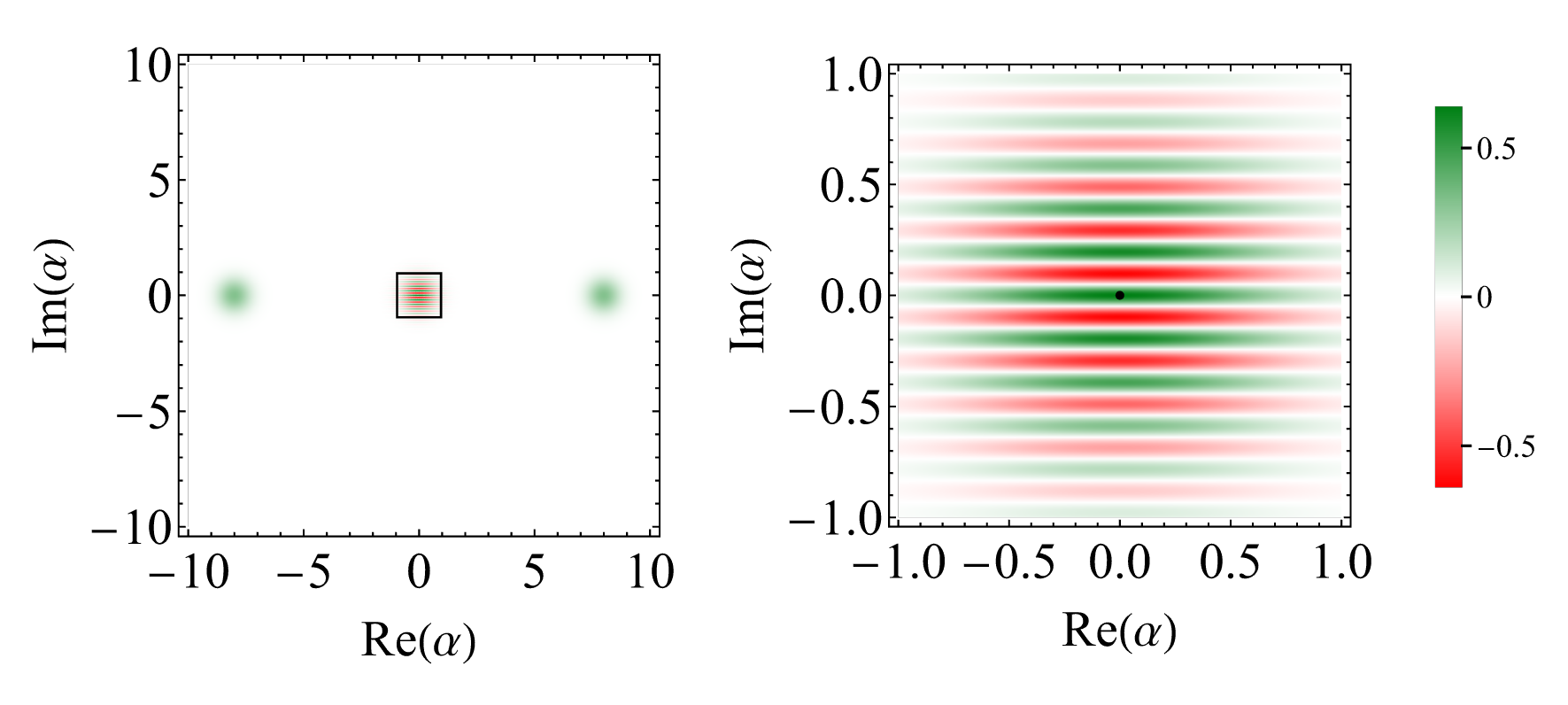}
\caption{Wigner distribution function for a two-component cat state (left panel) and the central interference region of the Wigner function (right panel). These intensity plots are generated with $\beta = 8$ and $\theta_1 = 0$ in Eq.~(\ref{eq:wignercat2}). The interference patch of interest is indicated by a black dot.}
\label{fig:cat2}
\end{figure}

\subsection{Cat versus compass state}

In this subsection, we examine two fundamental quantum states that exemplify the core concepts of our research. The first is a two-component cat state (commonly known as a ``cat state")~\cite{glauber_coherent_1963,dodonov_even_1974,yurke_generating_1986}, and the second is a four-component cat state, often called a ``compass state"~\cite{zurek_sub-planck_2001}. Figures \ref{fig:fig1}(a) and \ref{fig:fig1}(b) provide schematic representations of these two states in phase space. As shown, setting $L=2$ in the generic superposition of Eq.~(\ref{eq:genric1}) leads to a two-component or macroscopic cat state, which serves as a quintessential examples in our study~\cite {buzek_superpositions_1992,gerry_two-mode_1997,sanders_connection_2014,ELNASCHIE2005673,SACKETT2003431}. A two-component cat state is denoted as
\begin{equation}\label{eq:cat2}
\ket{\beta_2}= \frac{1}{\sqrt{2}} \Big(\ket{\beta}  + \mathrm{e}^{\mathrm{i}\theta_1}\ket{-\beta}\Big),
\end{equation}
where the phases with $\theta_1 = 0$ and $\theta_1 = \pi$ correspond to even and odd cat states~\cite{dodonov_even_1974}, respectively. For comparatively larger values of $\beta$, the superposition Eq.~(\ref{eq:cat2}) gives rise to properties such as superposition and entanglement, which make these superpositions useful for quantum computing~\cite{goto_bifurcation-based_2016,cochrane_macroscopically_1999} and communication~\cite{van_enk_entangled_2001,do_teleportation_2019,yin_coherent-state-based_2019}, to name a few applications. The Wigner function of the two-component cat state presented in Eq.~(\ref{eq:cat2}) yields
\begin{equation}\label{eq:wignercat2}
W_{2}\left( \alpha \right)= \frac{1}{\pi } \Big[ G\left( \beta \right) +  G\left( -\beta \right) +2 G\left( 0 \right)\cos{\left(\theta_1 + 4 \beta \text{Im}(\alpha)\right)}\Big],
\end{equation}
with $G\left( \Theta \right)= \mathrm{e}^{-2\left\lvert\alpha - \Theta\right\rvert^2}$ representing Gaussian distributions pinned at $\Theta$ in phase space. The Wigner function is represented in Fig.~\ref{fig:cat2} for $\theta_1=0$. Two Gaussian lobes, representing two-component coherent states, are fixed at specific locations in phase space, with the central oscillatory pattern indicating quantum interference. Our analysis concentrates on this interference pattern, which is of particular interest and is located around the phase-space origin.

\begin{figure}
\centering
\includegraphics[width=0.5\textwidth]{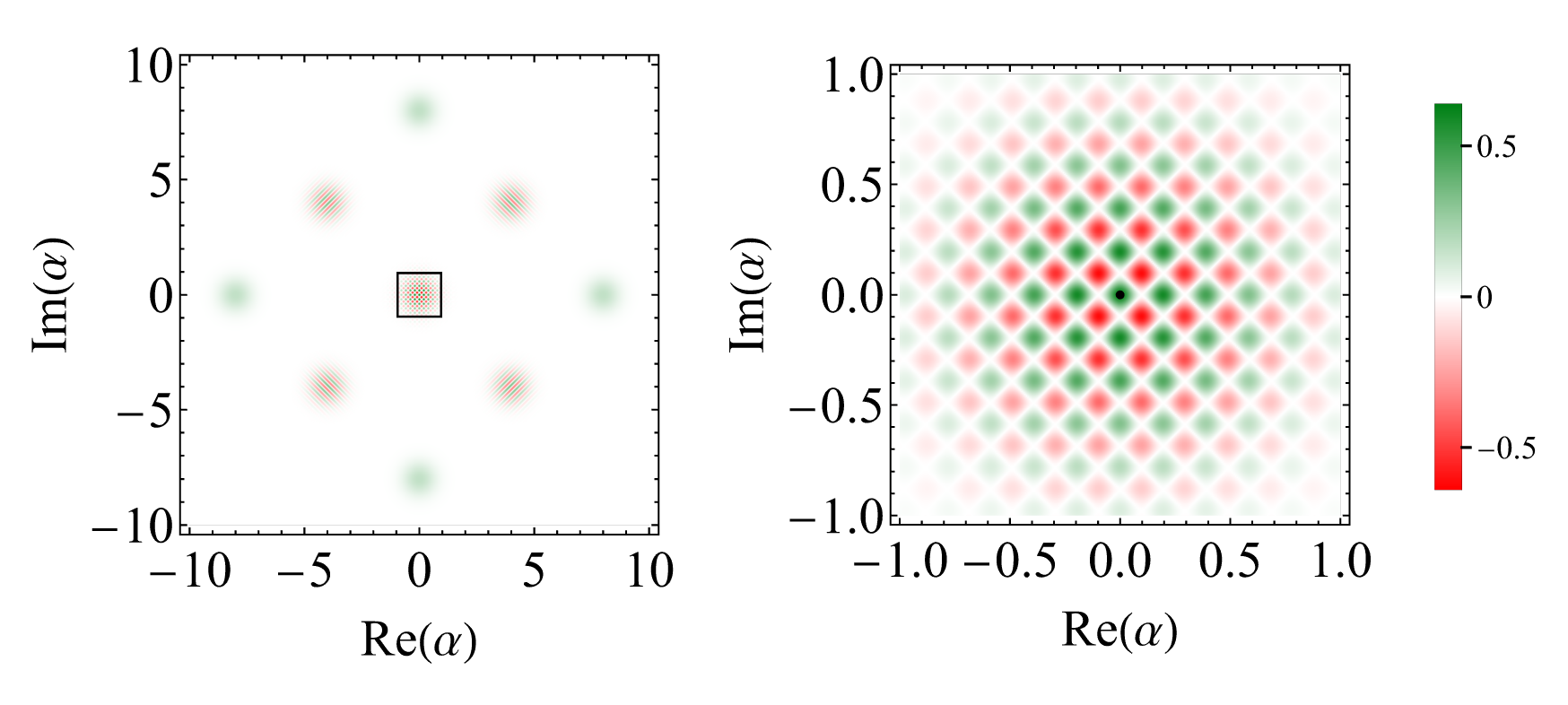}
\caption{Wigner distribution function for a four-component cat state (left panel) and the central sub-Planckian structures (right panel). These intensity plots were generated with 
$\beta = 8$ in Eq.~(\ref{eq:wignercat4}). The sub-Planckian structure of interest is indicated by a black dot.}
\label{fig:cat4}
\end{figure}

Consider the example of a four-component cat state, also known as a ``compass state"~\cite{zurek_sub-planck_2001}, obtained by setting $L=4$ in Eq.~(\ref{eq:genric1}). This quantum state was the first to introduce the concept of sub-Planckian structures in phase space. The sub-Planckian structures of this state are anisotropic, leading to an anisotropic enhancement in phase-space sensitivity \cite{shukla_superposing_2023}, which we discuss further. This compass state is denoted as
\begin{equation}\label{eq:cat4}
\ket{\beta_4}= \frac{1}{2} \Big(\ket{\beta}+ \ket{-\beta} + \ket{\mathrm{i}\beta} + \ket{-\mathrm{i}\beta}\Big),
\end{equation}
with the Wigner function
\begin{align}\label{eq:wignercat4}
W_{4}(\alpha)=&\nonumber \frac{1}{2\pi } \bigg[G( \beta) +  G( -\beta ) + G( \mathrm{i}\beta ) +  G( -\mathrm{i}\beta)\\
&\nonumber+ 2 G( \nicefrac{1+\mathrm{i}}{2})\cos{[-2 \beta \text{Re}(\alpha) - 2 \beta \text{Im}( \alpha) + \beta^2]}  \\
&\nonumber+2 G\left( \nicefrac{-1+\mathrm{i}}{2} \right)\cos{\left[2 \beta \text{Re}(\alpha)- 2 \beta \text{Im }(\alpha) + \beta^2\right]}\\
&\nonumber+2 G\left( \nicefrac{-1-\mathrm{i}}{2} \right)\cos{\left[2 \beta \text{Re}(\alpha) + 2 \beta \text{Im} (\alpha) + \beta^2\right]}\\
&\nonumber+2 G\left( \nicefrac{1-\mathrm{i}}{2} \right)\cos{\left[-2 \beta \text{Re}(\alpha) + 2 \beta \text{Im }(\alpha) + \beta^2\right]}\\
&+ 2 G\left(0\right)\cos{\left[4\beta \text{Im}(\alpha)\right]}+ 2 G\left(0\right)\cos{\left[4\beta \text{Re}(\alpha)\right]}\bigg].
\end{align}
Figure~(\ref{fig:cat4}) illustrates the Wigner function. The last two terms in Eq.~(\ref{eq:wignercat4}) generate a chessboardlike pattern in phase space, featuring anisotropic sub-Planckian structures that resemble tiles. By evaluating the zeros of the factor $ \cos{ \left(4\beta \text{Re}(\alpha)\right)} + \cos{ \left(4\beta \text{Im}(\alpha )\right)} $, one can determine the extension of the corresponding central patches, which in this case will be inversely proportional to the parameter $\beta$.

\begin{figure}[h]
\centering
\includegraphics[width=0.5\textwidth]{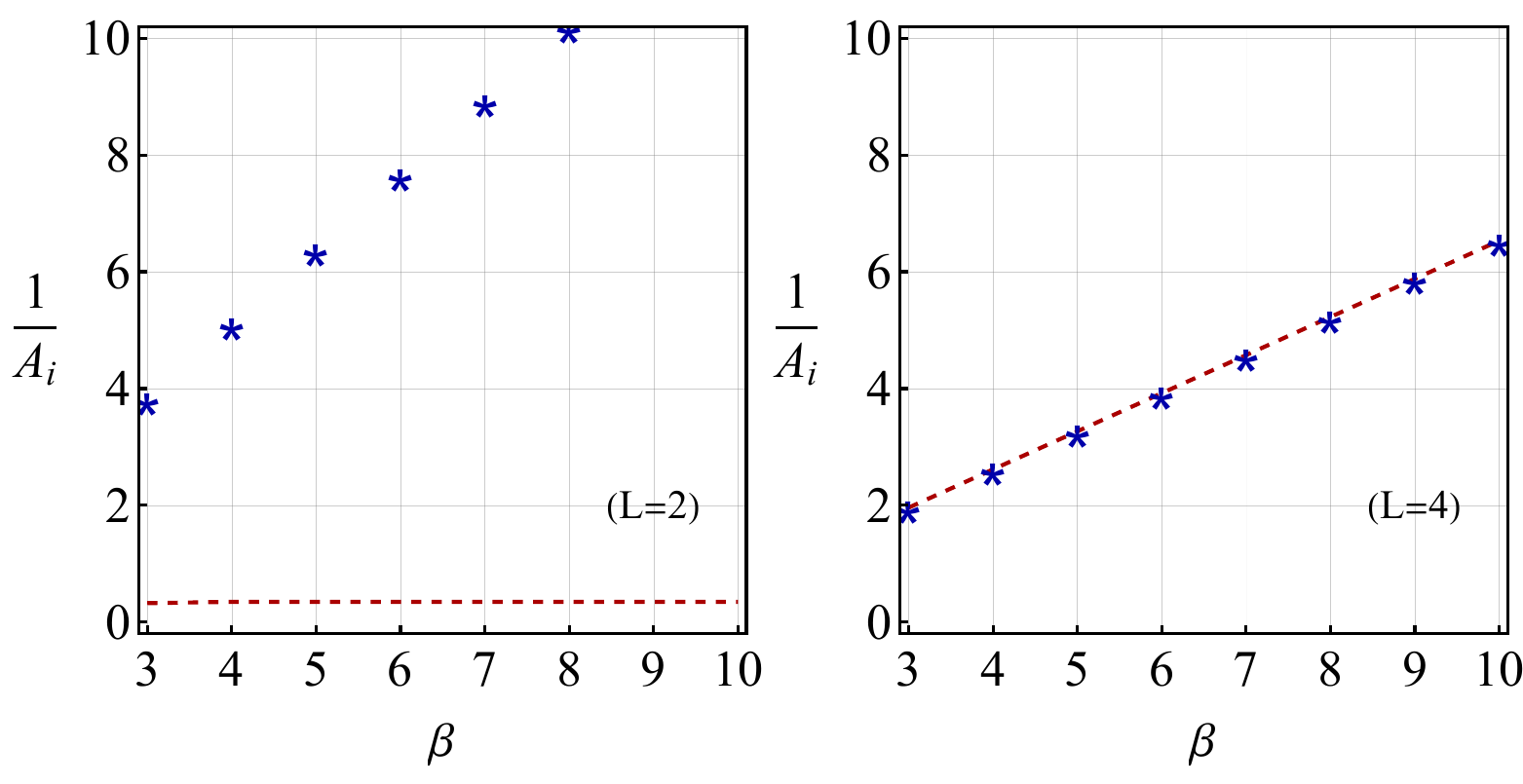}
\caption{Extensions of the center patch, defined by $A_i$, along the $\text{Re}(\alpha)$ and $\text{Im}(\alpha)$ directions in phase space. These correspond to Figs. \ref{fig:cat2} and \ref{fig:cat4}  for a two-component cat state ($L=2$) and a compass state ($L=4$), respectively. The pattern consisting of a sequence of stars indicates the $\text{Im}(\alpha)$ direction, while the red-dotted line represents the $\text{Re}(\alpha)$ direction.}
\label{fig:extensionsA}
\end{figure}

Let us now investigate the phase-space extensions along the $\text{Re}(\alpha )$ and $\text{Im}(\alpha )$ directions for the central interference patches in the cases of $L=2$ and $L=4$. We begin with the case of $L=2$. These extensions, indicated by $A_i$, are illustrated in Fig.~\ref{fig:extensionsA} for $L=2$. The central interference patch, marked by a black dot, has a constant extension along the $\text{Re}(\alpha)$ direction, with $A_i\sim1$, matching the extension as the coherent states is this specific direction in phase space. Conversely, the patch is limited by $A_i\sim \nicefrac{1}{\beta}$ along the $\text{Im}(\alpha)$ direction, indicating that for a two-component cat state, the central phase-space feature is limited in only one direction. 

In our study, we focus primarily on the central phase-space structure in each scenario, as the most intriguing phase-space features are located near the origin. For $L=4$, it is observed that a single patch exhibits extensions proportional to $\nicefrac{1}{\beta}$ along both the $\text{Re}(\alpha )$ direction and the $\text{Im}(\alpha )$ direction, thus being constrained in all phase-space directions. This is illustrated in Fig.~\ref{fig:cat4} for the case $L=4$, where the extensions $A_i$ are influenced by the parameter $\beta$. For $\beta \gg 1$, these extensions are significantly smaller than those of the coherent states, indicating the presence of the sub-Planckian structure.

In summary, the two-component cat state does not exhibit sub-Planckian structures in phase space. Conversely, the four-component cat state demonstrates anisotropic sub-Planckian structures. Specifically, when we assess the distance from the phase-space origin to the edges of a central feature, it becomes evident that this distance varies with the orientation. This variation highlights the anisotropic nature of the sub-Planckian structures in a four-component cat state. As discussed later, this anisotropic characteristic greatly impacts the sensitivity to displacements.

\subsection{Compass states with multiple components}

We now extend the concept of sub-Planckian structures to higher-order superposition multicomponent cat states, focusing on the cases shown in Figs.~\ref{fig:fig1}(c)–(f), obtained by setting $L = 6$, $L = 8$, $L= 10$, and $L=12$ in Eq.~(\ref{eq:genric1}). The cases with $L=8$ and $L=12$ are superpositions of two and three compass states, respectively, and were investigated recently to construct isotropic sub-Planckian features~\cite{shukla_superposing_2023}. We include these cases to compare them with our cases of 
$L=6$ and $L=10$, demonstrating that isotropic sub-Planckian structures are not limited to compass-state superpositions but can extend to certain $L$-component cat states, as our investigation confirms. This comparison shows that isotropic sub-Planckian structures from compass-state superpositions~\cite{shukla_superposing_2023} are a subset of the broader concept we propose.

\begin{figure}
\centering
\includegraphics[width=0.5\textwidth]{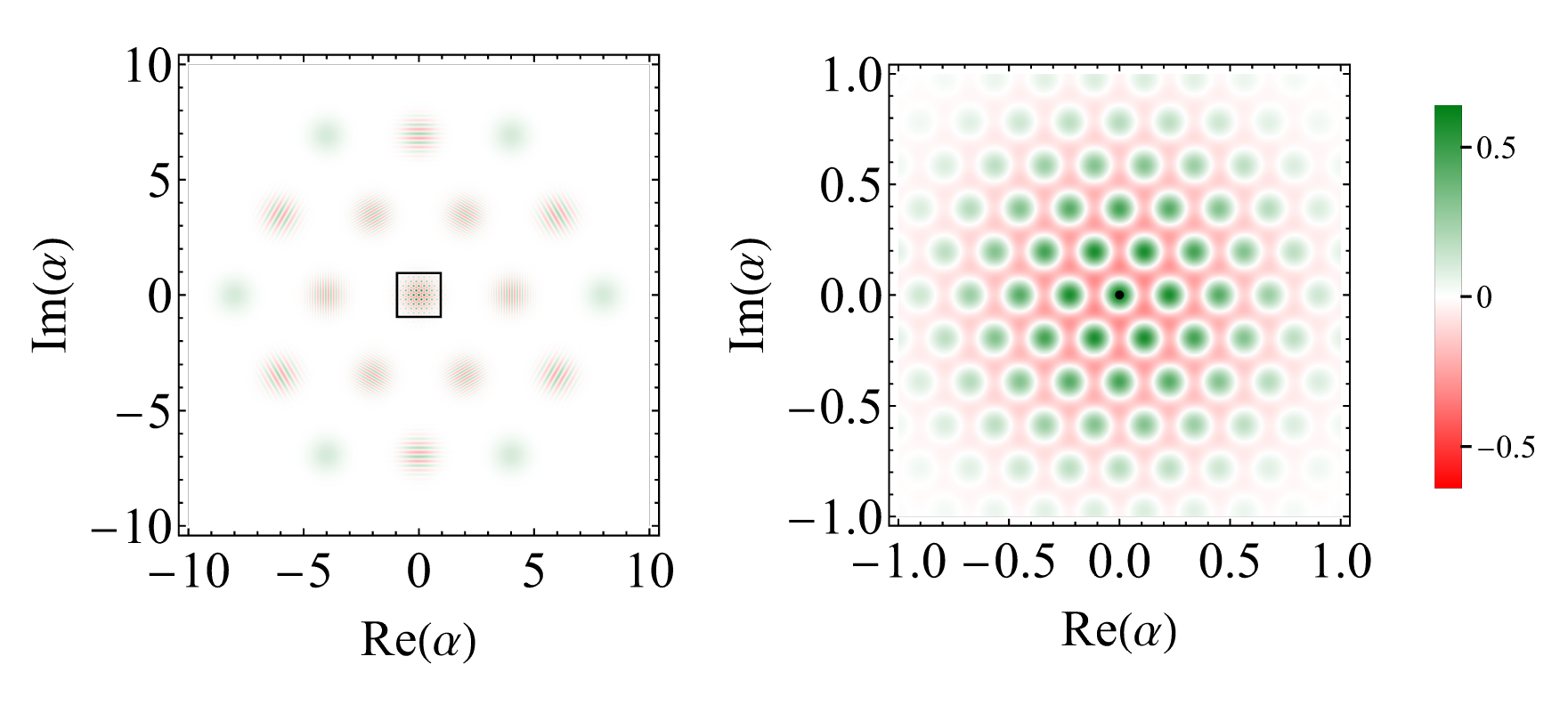}
\caption{Wigner distribution function for a six-component cat state (left panel) and corresponding central interference pattern (right panel), highlighting the isotropic central sub-Planckian structure with a black dot. These plots are generated with $L=6$ and $\beta = 8$ in Eq.~(\ref{eq:L_compo_Wigner}) with the appropriate $\omega_j$ values.}
\label{fig:cat6}
\end{figure}

First, we consider the six-component cat states with $L=6$:
\begin{align}
\ket{\beta_6}=&\nonumber \frac{1}{\sqrt{6}} \Big(
 \ket{\beta}+\ket{\mathrm{e}^{\mathrm{i}\nicefrac{\pi}{3}}\beta}  + \ket{\mathrm{e}^{\mathrm{i}\nicefrac{2\pi}{3}}\beta}+ \ket{-\beta} +\ket{\mathrm{e}^{-\mathrm{i}\nicefrac{2\pi}{3}}\beta} \\&+ \ket{\mathrm{e}^{-\mathrm{i}\nicefrac{\pi}{3}}\beta}\Big).
 \label{eq:cat6}
\end{align}
The corresponding Wigner function for this six-component cat state is shown in Fig.~\ref{fig:cat6}. The general form of the Wigner function is given in Eq.~(\ref{eq:L_compo_Wigner}), applicable here by setting $L=6$ and by using the relevant $\omega_j$ values. In line with the main objective of this work, we focus on the central interference part, denoted as $W_{\# L}\left(\alpha \right)$ for higher-order cases, and for this specific case, it takes the form
\begin{align}
W_{\# 6}\left( \alpha \right)=&\nonumber \frac{2G\left(0\right)}{3\pi}  \Big[ \cos{\left(4\beta \text{Im}(\alpha)\right)} +\cos( 2\sqrt{3} \beta \text{Re}(\alpha) \\ &- 2  \beta \text{Im}(\alpha))
+\cos(2\sqrt{3} \beta \text{Re}(\alpha) + 2  \beta \text{Im}(\alpha))
 \Big].
 \label{wignercat6}
\end{align}
As shown in Fig.~\ref{fig:cat6}, the central interference patch of this six-component cat state forms a circle, indicating that it is uniformly restricted in all phase-space directions, unlike a compass state. 

\begin{figure}[h]
\centering
\includegraphics[width=0.5\textwidth]{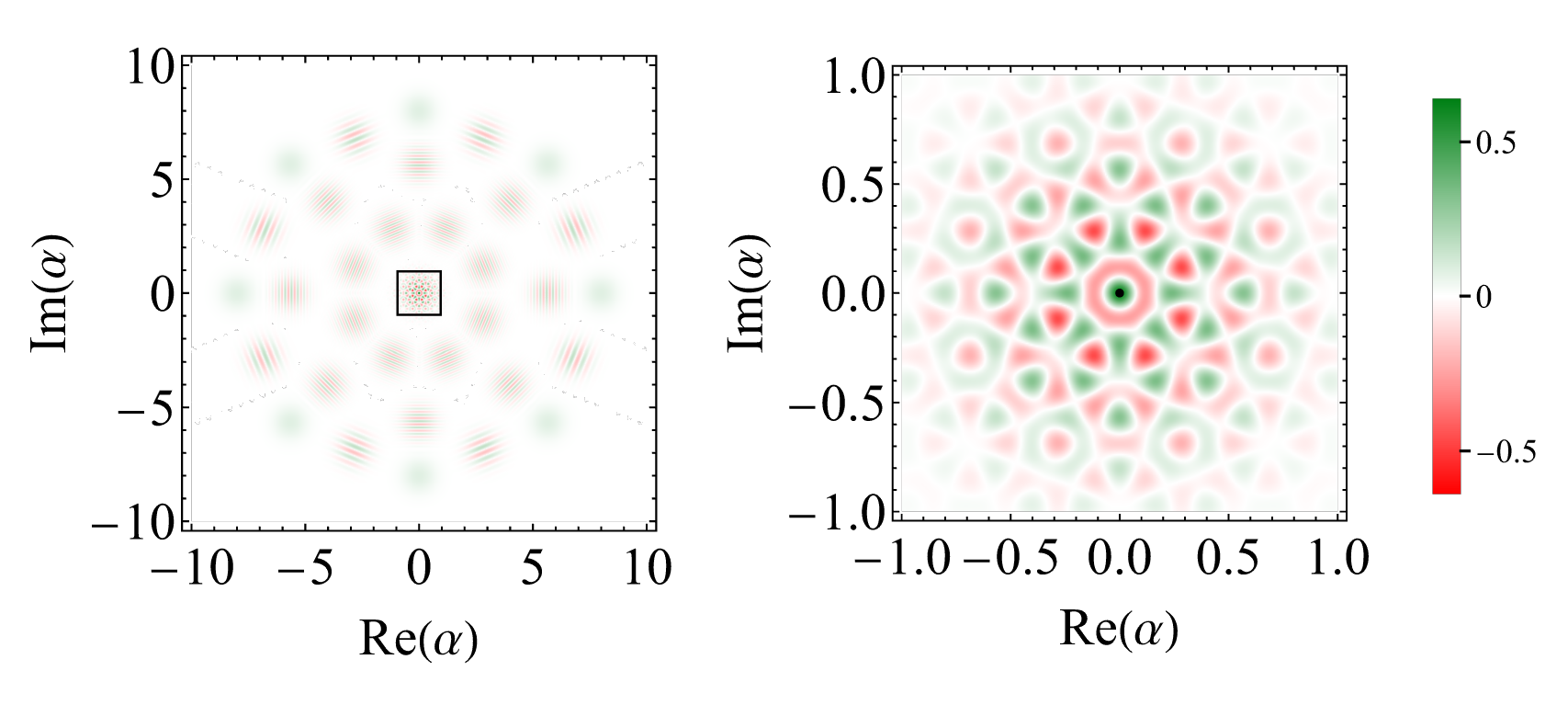}
\caption{Wigner distribution function for an eight-component cat state (left panel) and central interference pattern (right panel), with the patch marked by a black dot representing an isotropic sub-Planckian feature. These intensity plots are obtained with $L=8$ and $\beta = 8$ in Eq.~(\ref{eq:L_compo_Wigner}) with the corresponding $\omega_j$ values.}
\label{fig:cat8}
\end{figure}

\begin{figure}
\centering
\includegraphics[width=0.5\textwidth]{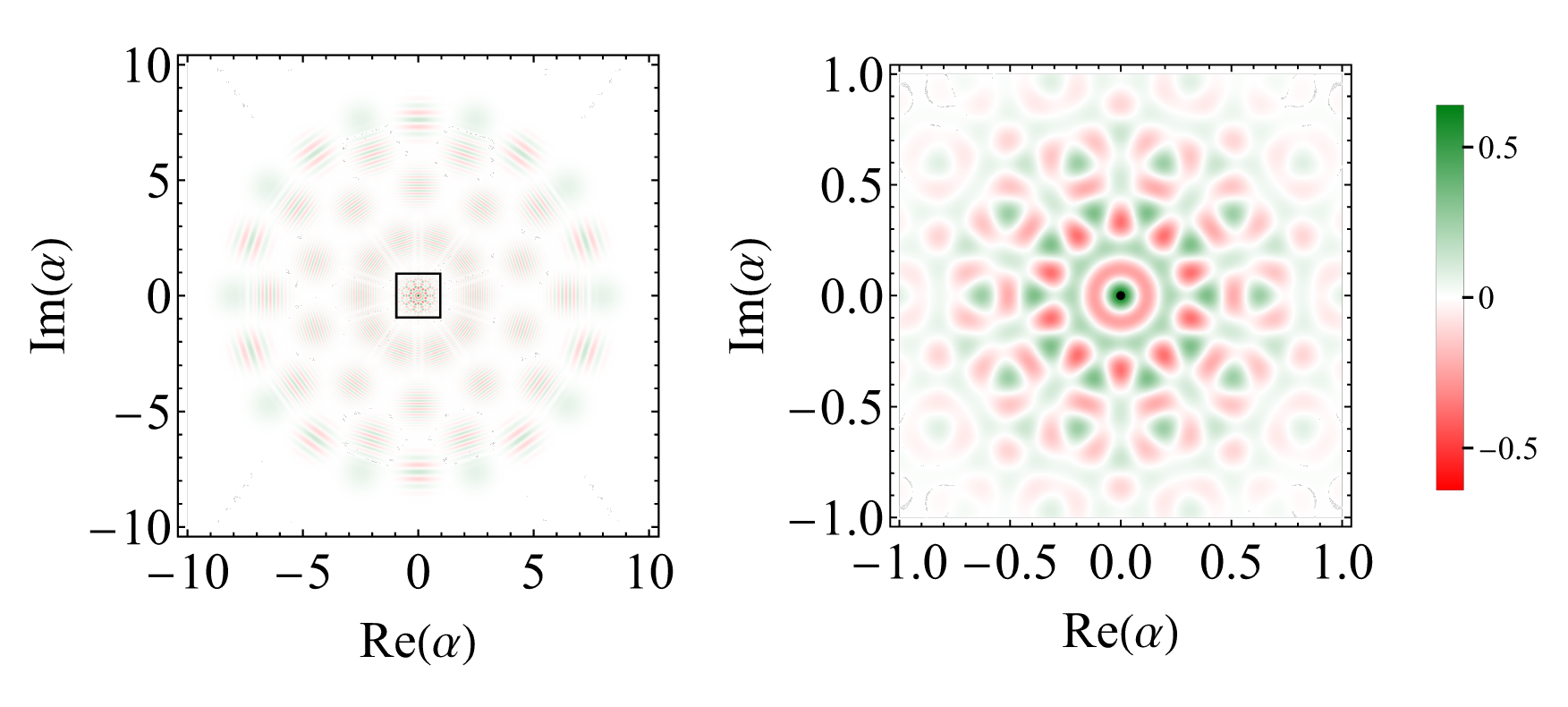}
\caption{Wigner distribution function for a ten-component cat state (left panel) and corresponding central cross section (right panel), with the isotropic sub-Planckian structure marked by a black dot. These plots are generated with $L=10$ and $\beta = 8$ in Eq.~(\ref{eq:L_compo_Wigner}) with adjusted $\omega_j$ values.}
\label{fig:cat10}
\end{figure}

Let us now explore the superposition of eight coherent states, also known as the superposition of four cat states or two compass states, as recently introduced in the context of isotropic sub-Planckian structures~\cite{shukla_superposing_2023}: 
\begin{align}
\ket{\beta_8}= &\nonumber \frac{1}{\sqrt{8}} \Big(
 \ket{\beta} + \ket{\mathrm{e}^{\mathrm{i}\nicefrac{\pi}{4}}\beta}  + \ket{\mathrm{i}\beta}  + \ket{\mathrm{e}^{\mathrm{i}\nicefrac{3\pi}{4}}\beta} + \ket{-\beta} \\
&  + \ket{\mathrm{e}^{-\mathrm{i}\nicefrac{3\pi}{4}}\beta}  + \ket{-\mathrm{i}\beta}  + \ket{\mathrm{e}^{-\mathrm{i}\nicefrac{\pi}{4}}\beta} \Big). 
\label{eq:cat8}
\end{align}
The Wigner function for this state is shown in Fig.~\ref{fig:cat8}, and is derived by adjusting the relevant parameters in Eq.~(\ref{eq:L_compo_Wigner}) with $L=8$. The central interference pattern exhibits isotropic sub-Planckian structures at the center of phase space, and this pattern can be mathematically expressed as
\begin{align}
&\nonumber W_{\# 8}\left( \alpha \right)=\\&\nonumber \frac{G\left(0\right)}{2 \pi} \Big[ \cos{\left(4\beta \text{Im}(\alpha)\right)} + \cos{\left(\sqrt{8}\beta (\text{Re}(\alpha)- \text{Im}(\alpha))\right)}\\& + \cos{\left(4\beta \text{Re}(\alpha)\right)} + \cos{\left(\sqrt{8}\beta (\text{Re}(\alpha)+ \text{Im}(\alpha))\right)} \Big].
\end{align}
Figure~\ref{fig:extensions} demonstrates the limitations of the isotropic sub-Planckian structure in phase space for $L=8$. A visual comparison with our $L=6$ case clearly shows the equivalence of central isotropic features, suggesting that an isotropic version of sub-Planckian structures can be achieved through the superposition of three cat states. This concept can be extended to the superposition of five cat states, referred to as ten-component cat states, as follows:
\begin{align}
    \ket{\beta_{10}}=&\nonumber \frac{1}{\sqrt{10}} \Big(\ket{\beta} + \ket{\mathrm{e}^{\mathrm{i}\nicefrac{\pi}{5}}\beta} + \ket{\mathrm{e}^{\mathrm{i}\nicefrac{2\pi}{5}}\beta} + \ket{\mathrm{e}^{\mathrm{i}\nicefrac{3\pi}{5}}\beta}\\&\nonumber+\ket{\mathrm{e}^{\mathrm{i}\nicefrac{4\pi}{5}}\beta}  + \ket{-\beta} + \ket{\mathrm{e}^{-\mathrm{i}\nicefrac{4\pi}{5}}\beta} + \ket{\mathrm{e}^{-\mathrm{i}\nicefrac{3\pi}{5}}\beta}\\&+\ket{\mathrm{e}^{-\mathrm{i}\nicefrac{2\pi}{5}}\beta} + \ket{\mathrm{e}^{-\mathrm{i}\nicefrac{\pi}{5}}\beta}\Big). 
    \label{eq:10_compo_cat}
\end{align}%
As previously discussed, the Wigner function for this state is obtained by setting $L=10$ with the relevant phases in Eq.~(\ref{eq:10_compo_cat}). Figure~\ref{fig:cat10} shows the corresponding Wigner function, illustrating central interference with isotropic sub-Planckian structures at the phase-space origin. The central interference in this case is given by
\begin{align}
   W_{\# 10}\left( \alpha \right)=  & \nonumber  \frac{2G\left(0\right)}{5 \pi} \Big[ \cos{\left(4\beta \text{Im}(\alpha)\right)} \\&+  \nonumber  \cos{\left(4\beta \cos{\frac{\pi}{5}}  \text{Im}(\alpha)  -  4\beta \sin{\frac{\pi}{5}} \text{Re}(\alpha)  \right)}\\&  +  
   \nonumber \cos{\left(4\beta \cos{\frac{2\pi}{5}}  \text{Im}(\alpha)  -  4\beta \sin{\frac{2\pi}{5}} \text{Re}(\alpha)  \right)} \\& +  \nonumber  
   \cos{\left(4\beta \cos{\frac{3\pi}{5}}  \text{Im}(\alpha)  -  4\beta \sin{\frac{3\pi}{5}} \text{Re}(\alpha)  \right)} \\& +  
   \cos{\left(4\beta \cos{\frac{4\pi}{5}}  \text{Im}(\alpha)  -  4\beta \sin{\frac{4\pi}{5}} \text{Re}(\alpha)  \right)} \Big] .  
\end{align}

\begin{figure}
\centering
\includegraphics[width=0.5\textwidth]{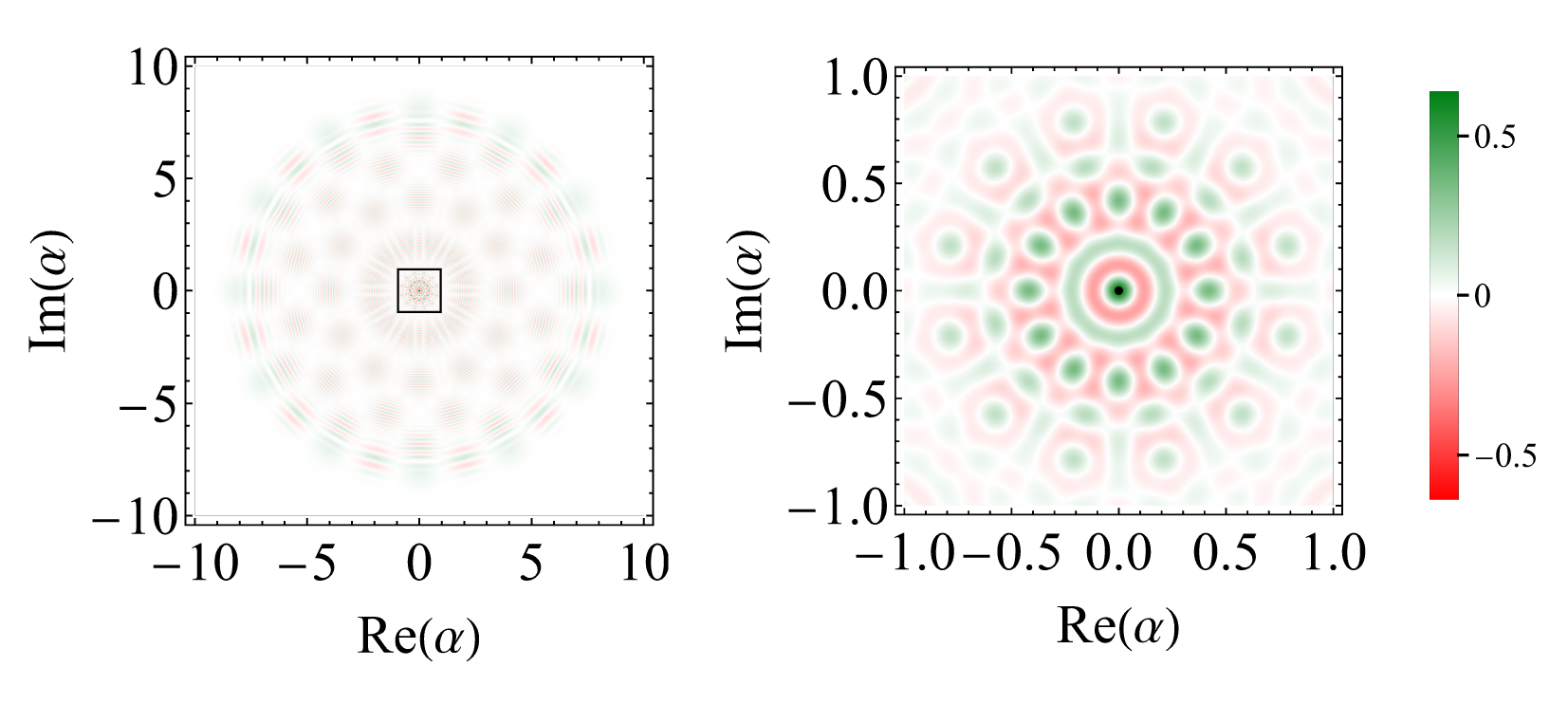}
\caption{Wigner distribution function for a 12-component cat state (left panel) and cross section of the central interference pattern (right panel), with the central isotropic sub-Planckian structure highlighted by a black spot. These intensity plots are generated with $L=12$ and $\beta = 8$ in Eq.~(\ref{eq:L_compo_Wigner}), with the $\omega_j$ values from the corresponding superposition.}
\label{fig:cat12}
\end{figure}

\begin{figure}
\centering
\includegraphics[width=0.5\textwidth]{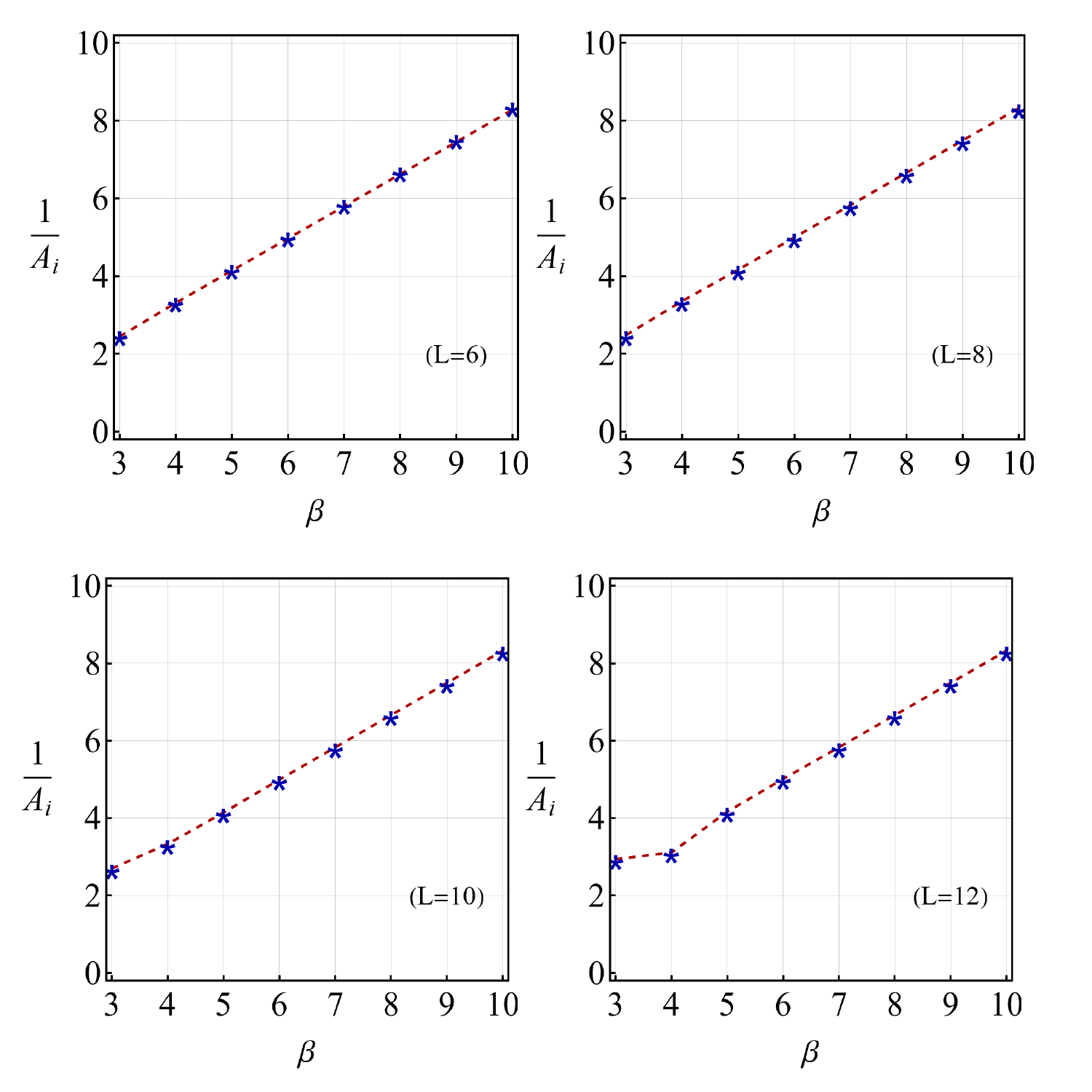}
\caption{Extensions of the central phase-space feature, denoted by $A_i$, along the $\text{Re}(\alpha)$ and $\text{Im}(\alpha)$ directions for the $L$-component cat states. Each example is labeled with its corresponding $L$ value. The pattern consisting of a sequence of stars indicates the $\text{Im}(\alpha)$ direction, while the red-dotted line represents the $\text{Re}(\alpha)$ direction in phase space.}
\label{fig:extensions}
\end{figure}

Finally, the 12-component cat state, essentially a superposition of three compass states as previously discussed~\cite{shukla_superposing_2023}, presents a situation similar to our cases. As shown in Fig.~\ref{fig:cat12}, this specific case is denoted by
\begin{align}
   &\nonumber \ket{\beta_{12}}=\\&\nonumber \frac{1}{\sqrt{12}} \Big(\ket{\beta} + \ket{\mathrm{e}^{\mathrm{i}\nicefrac{\pi}{6}}\beta} + \ket{\mathrm{e}^{\mathrm{i}\nicefrac{\pi}{3}}\beta} + \ket{\mathrm{i}\beta}+\ket{\mathrm{e}^{\mathrm{i}\nicefrac{2\pi}{3}}\beta}\\&\nonumber +\ket{\mathrm{e}^{\mathrm{i}\nicefrac{5\pi}{6}}\beta} + \ket{-\beta} + \ket{\mathrm{e}^{-\mathrm{i}\nicefrac{5\pi}{6}}\beta}+\ket{\mathrm{e}^{-\mathrm{i}\nicefrac{2\pi}{3}}\beta}+  \ket{-\mathrm{i}\beta}  \\&+ \ket{\mathrm{e}^{-\mathrm{i}\nicefrac{\pi}{3}}\beta} + \ket{\mathrm{e}^{-\mathrm{i}\nicefrac{\pi}{6}}\beta}\Big), 
    \label{eq:12_compo_cat}
\end{align}
with the Wigner function as given by Eq.~(\ref{eq:L_compo_Wigner}) and exhibiting central interference denoted by 
\begin{align}
W_{\# 12}\left( \alpha \right)= & \nonumber  \frac{G\left(0\right)}{3 \pi} \Big[ \cos{\left(4\beta \text{Re}(\alpha)\right)}+\cos{\left(4\beta \text{Im}(\alpha)\right)}\\& \nonumber + \cos{\left(2\beta \text{Re}(\alpha) - 2 \sqrt{3}  \beta \text{Im}(\alpha)\right)}\\& \nonumber + \cos{\left(2\sqrt{3}\beta \text{Re}(\alpha) - 2  \beta \text{Im}(\alpha) \right)}
\\& \nonumber + \cos{\left(2\sqrt{3}\beta \text{Re}(\alpha) + 2  \beta \text{Im}(\alpha) \right)} \\&  + \cos{\left(2\beta \text{Re}(\alpha) + 2  \sqrt{3}\beta \text{Im}(\alpha) \right)}
\Big],
\end{align}
which can be generalized to
\begin{align}
   &\nonumber W_{\# L }\left( \alpha \right)= \\& \frac{4 G\left(0\right)}{\pi L} \sum_{j=0}^{\nicefrac{L}{2}-1} \cos{\left(4\beta \cos{\omega_j}  \text{Im}(\alpha)  - 4\beta \sin{\omega_j}  \text{Re}(\alpha) \right)}.
\end{align}
 This central pattern encloses isotropic sub-Planckian structures at the phase-space origin, as shown in Fig.~\ref{fig:cat12}.

Figure~\ref{fig:extensions} presents the extension of central isotropic phase-space structures for 
 $L = 6$, $L = 8$, $L= 10$, and $L=12$ as a function of the parameter $\beta$. Each case is labeled with its corresponding 
$L$ value, demonstrating that the area of the central feature is significantly smaller than that of coherent states, thus maintaining a sub-Planck-scale in phase space. This section includes examples of superpositions involving six and ten coherent states, as well as previously studied compass-state superpositions, which may exhibit similar isotropic sub-Planckian features~\cite{shukla_superposing_2023}. These findings suggest that the analyzed cases also display isotropic sub-Planckian structures, with compass-state superpositions serving as a specific instance of our broader illustration. Furthermore, we demonstrate that these examples may exhibit enhanced isotropic sensitivity to displacements.

\begin{figure*}
\centering
\includegraphics[width=0.9\textwidth]{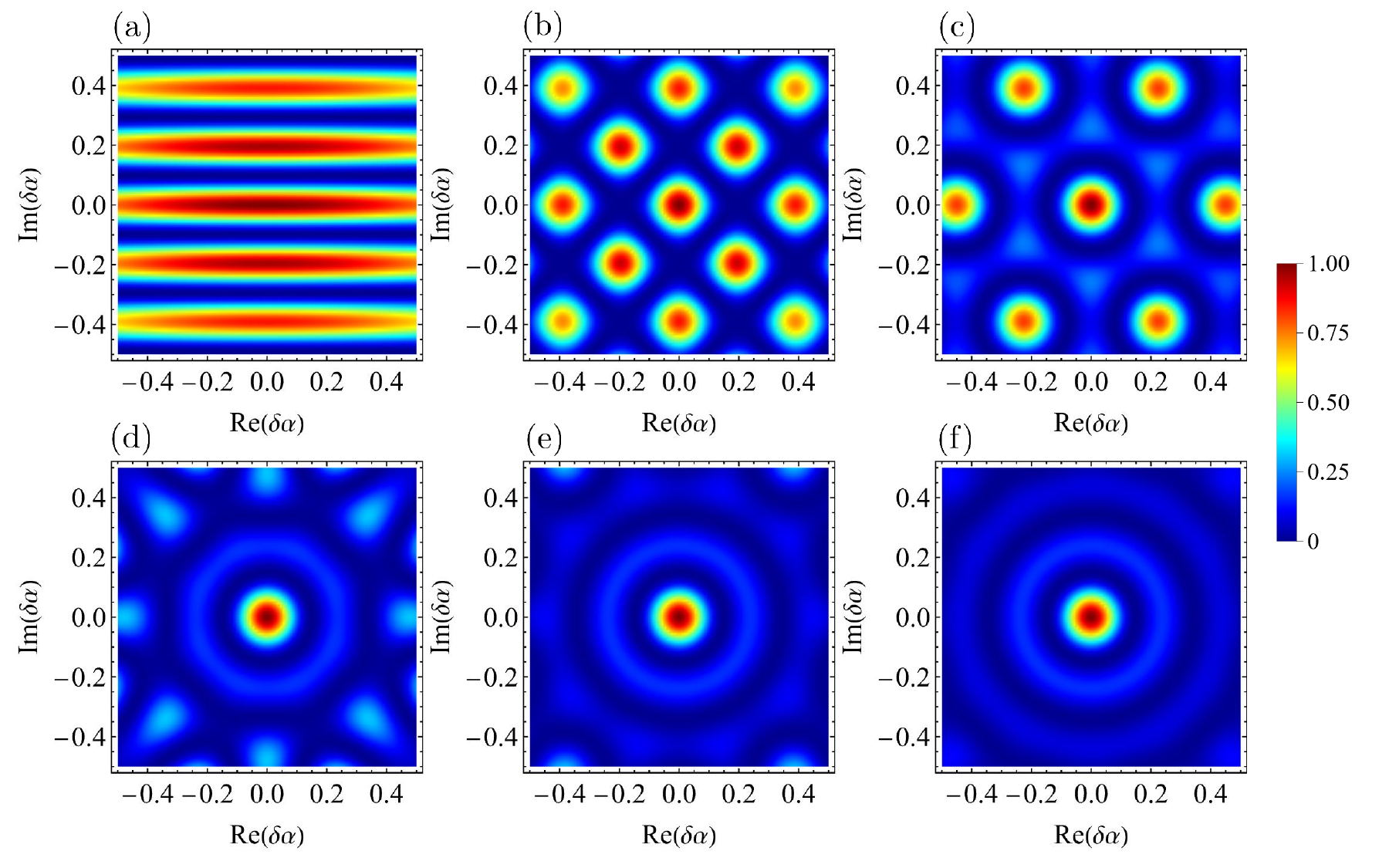}
\caption{Overlap function of $L$-component cat states: (a) $L=2$, (b) $L=4$, (c) $L=6$, (d) $L=8$, (e) $L=10$, and (f) $L=12$. In all cases $\beta=8$.}
\label{fig:overlaps}
\end{figure*}

\section{ENHANCEMENT IN SENSITIVITY}\label{subsec:sensitivity1}

In this section, we first address the sensitivity to displacements of the quantum states presented in the preceding section, and then we highlight their relevance in quantum measurements, particularly in the context of weak-force detection.

\subsection{Isotropically enhanced sensitivity}

External perturbations, such as displacements, can alter a quantum state in phase space, potentially making the displaced quantum state orthogonal to its original form. The phase-space characteristics of a quantum state are intrinsically linked to its sensitivity to such displacements~\cite{toscano_sub-planck_2006}. This relationship can be examined by evaluation of the overlap between a quantum state 
$\hat{\rho}$ and its slightly displaced counterpart $\hat{\rho}_{d} = \hat{D}\left(\delta\alpha\right)\hat{\rho}\hat{D}^\dagger\left(\delta\alpha\right)$, where $\delta \alpha = \delta x + \mathrm{i} \delta p$~\cite{akhtar_sub-planck_2021}. Here $\delta x$ and $\delta p$ represent the displacement amounts along the $\text{Re}(\alpha)$ and $\text{Im}(\alpha)$ directions in phase space, respectively. This overlap serves as a measure of the sensitivity to phase-space displacements for a quantum state. The overlap between a quantum state and its $\delta \alpha$-displaced counterpart is given by
\begin{equation}\label{eq:overlap}
O\left(\delta x , \delta p\right) := \mathrm{Tr} \left(\hat{\rho}\hat{\rho}_{d} \right) = \pi \int_{\inf} W_{\hat{\rho}}\left(\alpha\right)  W_{\hat{\rho}} \left(\alpha+\delta \alpha \right)  \mathrm{d}^2 \alpha.
\end{equation}
This relationship shows that sensitivity to displacement is proportional to the finer scale available in the phase space for quantum states. That is, states with smaller features exhibit greater sensitivity. For instance, compass states possess sub-Planckian phase-space characteristics, leading to sensitivity beyond the typical Planckian limits~\cite{zurek_sub-planck_2001}.

A slightly displaced $L$-component cat state is denoted as
\begin{equation}
\hat{D}\left( \delta \alpha\right)\ket{\beta_L}=\frac{1}{\sqrt{L}} \sum_{j=0}^{L-1} \mathrm{e}^{\mathrm{i}\left(-\beta \delta x \sin{\omega_j}  +  \beta \delta p \cos{\omega_j} \right)}\ket{\beta \mathrm{e}^{\mathrm{i}\omega_j}+\delta \alpha},
\end{equation}
where $\omega_j = \nicefrac{2 \pi j}{L}$, and the overlap function is expressed as
\begin{align}
&\nonumber O\left(\delta x,\delta p\right) = \left\lvert\braket{\beta_L \mid \hat{D}\left( \delta \alpha\right) \mid \beta_L } \right\rvert^2= \frac{\mathrm{e}^{ - 2\beta^2 - |\delta \alpha|^2}}{L^2}\\\times
&\left\lvert \sum_{j,k}^{L-1} \exp{\left[ \beta^2  \mathrm{e}^{\mathrm{i}(\omega_j-\omega_k)}- \beta \delta \alpha^*\mathrm{e}^{+\mathrm{i}\omega_j } +  \beta \delta \alpha\mathrm{e}^{-\mathrm{i}\omega_k}   \right] }\right\rvert ^2.
\end{align}
In the limit where $\beta \gg 1$, only the terms with $j=k$ in the summation contribute significantly. Therefore, the overlap function can be expressed as
\begin{equation}\label{eq:overlap_3}
O\left(\delta x,\delta p\right) \approx \frac{\mathrm{e}^{ - |\delta \alpha|^2}}{L^2}\left\lvert\sum_{j}^{L-1}\mathrm{e}^{2\mathrm{i}\beta\left( \delta p \cos{\omega_j} - \delta x \sin{\omega_j}\right)}\right\rvert ^2.
\end{equation}
The sensitivity of a quantum state to displacements is characterized by  the minimal change in displacement $\delta \alpha$ needed to reduce the overlap between states to zero. For a coherent state $\ket{\beta}$, the overlap between its $\delta \alpha$-displaced version is given by $O(\delta \alpha) = \text{e}^{-|\delta\alpha|^2}$, indicating that the smallest detectable displacement must exceed the Planck scale, specifically $|\delta\alpha| > 1$.

\begin{figure}
\centering
\includegraphics[width=0.51\textwidth]{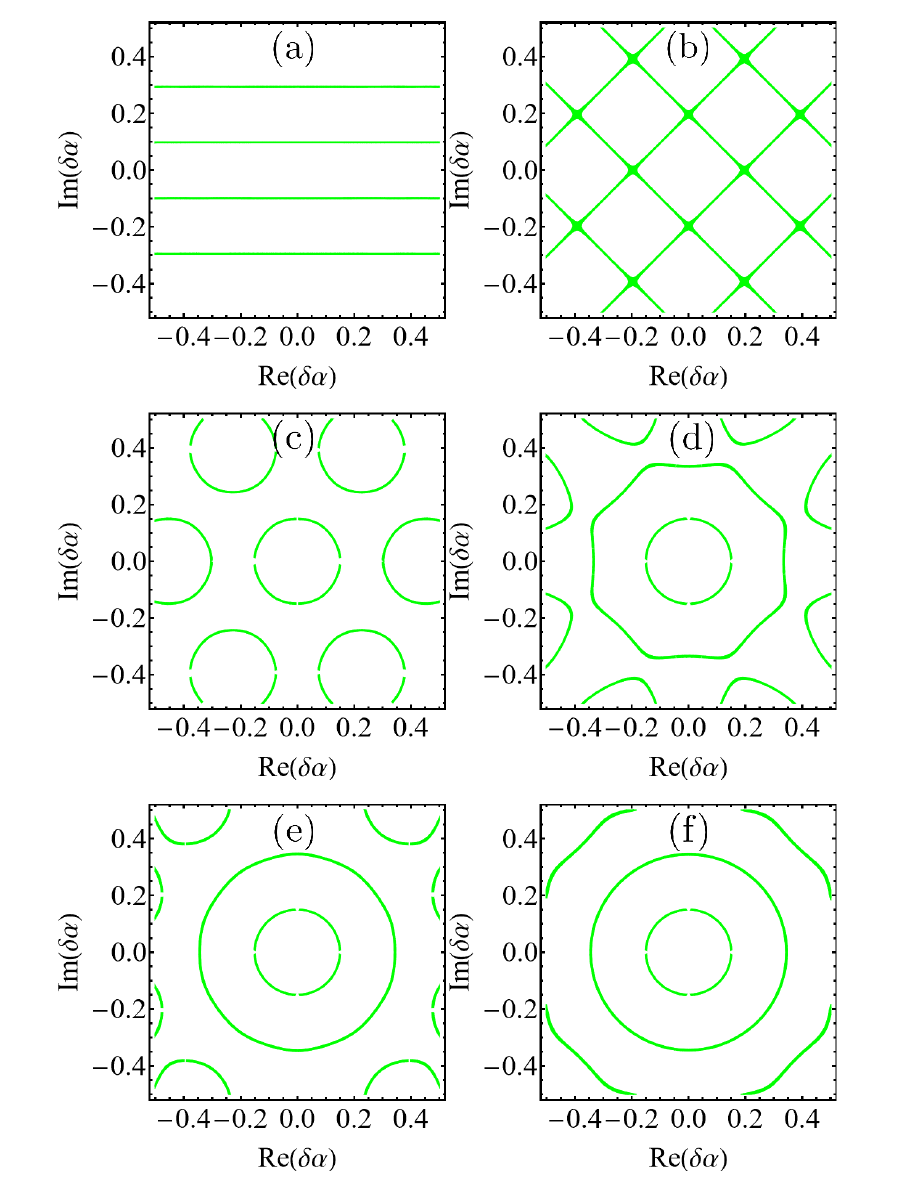}
\caption{Regions where the overlap function is approximately zero (less than $10^{-4}$): (a) $L=2$, (b) $L=4$, (c) $L=6$, (d) $L=8$, (e) $L=10$, and (f) $L=12$. We set $\beta=8$ in all cases.}
\label{fig:zeros}
\end{figure}

First, we consider the two-component cat state ($L=2$). Figure~\ref{fig:overlaps}(a) illustrates the overlap function for this example. In this context, we highlight the regions where the overlap function exhibits minor amplitudes, as illustrated in Fig.~\ref{fig:zeros}(a). This emphasizes the extent of displacement required for the cat state and its slightly displaced variant to become orthogonal. It can be demonstrated that the core region contributes to the overlap, with its corresponding zeros retaining their shape during variations of $\beta$ along the $\delta x$ direction, while they shrink along the $\delta p$ direction. The overlap becomes zero for a displacement $\delta p\approx \nicefrac{1}{\beta}$ along the $\text{Im}(\alpha)$ direction, while a constant displacement $\delta x >1$ is required along the $\text{Re}(\alpha)$ direction to achieve the vanishing of this overlap. This is identical to the behavior of coherent states, where the orthogonality of the cat state in this direction occurs at the Planck-scale level of displacement. This indicates that sensitivity to displacement for this case is enhanced compared with that for a coherent state along the $\text{Im}(\alpha)$ direction, where it is less than that of coherent states, but along the $\text{Re}(\alpha)$ direction, it matches that of coherent states. Thus, for an ordinary cat state, enhanced sensitivity is directed along a specific phase-space direction~\cite{akhtar_sub-planck_2021}.

We now examine the overlap function for states with sub-Planckian phase-space structures, focusing on the four-component cat state ($L=4$). This scenario is depicted in Fig.~\ref{fig:overlaps}(b), with the corresponding zeros shown in Fig.~\ref{fig:zeros}(b). It is observed that displacement $\delta x\approx \nicefrac{1}{\beta}$ along the $\text{Re}(\alpha)$ direction and $\delta p\approx  \nicefrac{1}{\beta}$ along the $\text{Im}(\alpha)$ direction is required to nullify the overlap. This is also true for other phase-space directions; specifically, as $\beta$ increases, the central feature of the overlap shrinks in all directions simultaneously. However, as shown in Fig.~\ref{fig:zeros}(b), the overlap vanishes anisotropically, confined to anisotropic regions. This indicates that in compass states, sensitivity enhancement is anisotropic, meaning that the anisotropic sub-Planckian structures of the compass state lead to anisotropic sensitivity to displacements.

In contrast to cat states and compass states, as illustrated in Figs.~\ref{fig:overlaps}(c)–\ref{fig:overlaps}(f) the overlaps of multicomponent cat states ($L\geq6$) with isotropic sub-Planckian structures vanish in isotropic regions. Figure~\ref{fig:overlaps}(c) shows the six-component cat state, with overlap zeros in Fig.~\ref{fig:zeros}(c). The eight-component cat-state overlap is shown in Fig.~\ref{fig:overlaps}(d), with zeros in Fig.~\ref{fig:zeros}(d). Figures ~\ref{fig:overlaps}(e) and \ref{fig:overlaps}(f) illustrate the ten-component and 12-component cat states, with their zeros in Figs.~\ref{fig:zeros}(e) and \ref{fig:zeros}(f). In all cases examined, a circular pattern centered at the origin is observed in both overlaps, which is further confirmed by their corresponding zeros, indicating isotropic sensitivity enhancement. Note that for the cases when $L\gg 1$, the dependence of the corresponding overlap function on $L$ is negligible. For example, for the cases $L\geq6$ as shown in Figs.~\ref{fig:overlaps}(c)–\ref{fig:overlaps}(f), the structures appearing at the center are similar, and this similarity becomes more pronounced as the number of coherent states in the specified superpositions increases. In the case when $L\to \infty$ Eq.~(\ref{eq:overlap_3}) can also be represented in terms of the first kind of Bessel function $J_n(x)$; that is,
\begin{align}
O\left(\delta x,\delta p\right)\approx \mathrm{e}^{- |\delta \alpha|^2} \left\lvert J_0\left( 2 \beta |\delta \alpha|\right)\right\rvert^2.
\end{align}
This indicates that the isotropic enhancement in sensitivity is driven primarily by the superpositions of six coherent states, and as the number of coherent states increases in even increments, the corresponding sensitivities reach a constant value at sufficiently large $L$.

\subsection{Implications for the accuracy of measurements}

In the previous section, we analyzed the overlap between quantum states and their slightly-$\delta\alpha$-perturbed versions, where quantum states with finer phase-space features were found to become orthogonal to their original (nonperturbed) versions at smaller $\delta\alpha$ values. This quantifies the capacity of corresponding states to detect minor perturbations, with states becoming distinguishable at smaller displacements being more sensitive. Here we provide a more insightful discussion of this concept.

Nonclassical states are the aspirations of enhanced quantum measurements~\cite{deng2024quantum,toscano_sub-planck_2006,dalvit_quantum_2006,Kolar2024negativewigner,PhysRevA.45.5193,Schneider1997zi}. For example, in the detection of weak forces, quantum states with finer phase-space features have been found to be more effective~\cite{toscano_sub-planck_2006,dalvit_quantum_2006}. The accuracy of parameter estimation is influenced by the energy resources (mean photon number $\bar{n}$) used. Minimal uncertainty states, such as coherent states, display sensitivity at the standard quantum limit, where, for example, as observed in the previous section, the sensitivity to displacement does not depend on $\bar{n}$. That is, in the standard dimensionless phase space, for a coherent state with an average photon number $\bar{n}$, a displacement $|\delta \alpha| \sim 1$ is required to achieve orthogonality between the original coherent state and its $\delta \alpha$-displaced version, implying that increasing $\bar{n}$ does not increase sensitivity; instead, the sensitivity is fundamentally constrained by \textit{shot noise} due to vacuum fluctuations.

The two-component cat state [Eq.~(\ref{eq:cat2})], shown in Fig.~\ref{fig:cat2}, is formed by superposing two displaced copies of coherent states along the position axis. This results in finer fringes along the momentum axis as the separation between the component coherent states increases (when $\beta \gg 1$). Notably, the central interference patch exhibits reduced uncertainty along the momentum quadrature, decreasing in proportion to $\nicefrac{1}{\beta}$ compared with the coherent-state level. As illustrated in Figs.~\ref{fig:overlaps}(a) and \ref{fig:zeros}(a) for the case when $\beta\gg1$, a cat state with average photon numbers $\bar{n}\propto|\beta|^2$ exhibits enhanced sensitivity compared with a coherent state. This enhancement occurs along momentum quadratures, scaling as $\nicefrac{1}{\sqrt{\bar{n}}}$. For $\beta\gg1$, this sensitivity exceeds the Planck scale, often referred to as the \textit{Heisenberg limit}, indicating that a cat state has greater sensitivity than coherent states with the same average photon number $\bar{n}$. 

The superposition of two cat states distributed along the position and momentum axes [the compass or four-component cat state from Eq.~(\ref{eq:cat4})] leads to the central phase-space feature whose dimensions compared with those of the coherent state are confined along all phase-space directions simultaneously. These central interference fringes (Fig.~\ref{fig:cat4}) reduce uncertainty in the position and momentum quadratures, resulting in sub-Planckian structures. In this case, the central superoscillatory pattern is limited by $\nicefrac{1}{\beta}$ along arbitrary directions in phase space. Compared with coherent states and two-component cat states, this multicomponent cat state exhibits enhanced sensitivity to displacement, scaling as $\nicefrac{1}{\sqrt{\bar{n}}}$. However, this enhancement is direction independent in phase space (Fig.~\ref{fig:overlaps}), highlighting the superiority of the multicomponent cat state over two-component cat states.

Two-component cat states and four-component cat states has been used to detect weak forces~\cite{toscano_sub-planck_2006,dalvit_quantum_2006}. It has been evidenced that quantum states with coarser phase-space structures, such as two-component cat states, are less effective than those with finer phase-space features, such as sub-Planckian structures. Specifically, in the case of two-component cat states, the detection of minor displacements diminishes when the direction of the perturbing force deviates from the axis perpendicular to the line connecting the two coherent states in superposition, while four-component cat states have the capacity to detect minor displacements of any direction in the phase space. This allows the simultaneous measurement of conjugate variables (position and momentum) associated with displacement with maximum precision, as shown in Fig.~\ref{fig:overlaps}. However, the anisotropic sub-Planckian structures observed in the compass state lead to an anisotropic enhancement of sensitivity to displacements, as illustrated in Fig.~\ref{fig:zeros}(b). 
Specifically, for the four-component cat state (or ordinary compass state), this enhancement in sensitivity exceeds that of the previous two-component cat state, although it remains direction dependent (anisotropic regions). Certain directions exhibit greater sensitivity enhancements than others, as indicated in Fig.~\ref{fig:zeros}(b). 
In contrast to the anisotropic sub-Planckian features, the isotropic counterparts identified in our cat states, which comprise six or more components, demonstrate a more favorable isotropic sensitivity enhancement. This is illustrated in Figs.~\ref{fig:cat6}–\ref{fig:cat12}, where isotropic sensitivity enhancement are shown, further verified in Figs.~\ref{fig:overlaps}(c)–\ref{fig:overlaps}(f) and Figs.~\ref{fig:zeros}(c)–\ref{fig:zeros}(f). In other words, the sensitivity of the six- or higher-component cat states we introduced is increased to the same scale ($\nicefrac{1}{\sqrt{\bar{n}}}$) as that of the four- or two-component cat states. However, this amplification occurs isotropically, which may have a substantial impact on the effectiveness of measurement techniques. Recent research~\cite{deng2024quantum} has investigated the fineness of phase-space properties of states and their impact on measurements. In our situation, it is apparent that the $L$-component cat states of specified shapes increase both the fineness of phase-space features and the sensitivity. This implies that, for the same amount of excitations $\bar{n}$, our higher-order $L$-component cat states might offer a more beneficial alternative for improving quantum measurements.

\section{Optomechanical modeling}\label{sec:optomechanical}

We present two scenarios related to the optomechanical setup: one with zero decoherence effects and another that accounts for these effects.

\subsection{Weak decoherence}

In the previous section, we detailed the unique phase-space characteristics of $L$-component cat states. Building upon this, we now explore their realization in an optomechanical framework, providing analogues  within this context~\cite{bose_preparation_1997}. This extension allows us to examine the physical manifestation and properties of these quantum states in a tangible system. Consider a system with a single-mode cavity field $\hbar \omega_{\mathrm{c}} \hat{a}^\dagger\hat{a}$ of frequency $\omega_{\mathrm{c}}$ with a movable mirror $\hbar \omega_{\mathrm{m}} \hat{b}^\dagger\hat{b}$ of frequency $\omega_{\mathrm{m}}$, acting as a harmonic oscillator. Here the radiation pressure force serves as a small perturbation on the mechanical motion of the mirror~\cite{law_interaction_1995}. In this scenario, the total Hamiltonian is given as
\begin{equation} \label{hami}
		\hat{H} = \omega_{\mathrm{c}}\hat{a}^\dagger \hat{a} + \omega_{\mathrm{m}} \hat{b}^\dagger \hat{b} - k \hat{a}^\dagger \hat{a} (\hat{b}^\dagger + \hat{b}),
\end{equation}
where $k$ is the optomechanical coupling constant, which quantifies the displacement of the zero-point motion of the mechanical oscillator caused by the momentum of a single photon. The Hamiltonian reveals the nonlinear coupling between the cavity field and the moving mirror, as the interaction term involves a product of three field operators. In the following analysis, we set the scaled frequency $\omega_{\mathrm{m}} = 1$.

In a typical setup, the initial state is a direct product of the cavity field and mirror states, denoted as $\ket{\Psi(0)}= \ket{\psi}_{\mathrm{c}} \otimes \ket{\phi}_{\mathrm{m}} $. Using the evolutionary operator decomposition from Refs.~\cite{mancini_ponderomotive_1997,bose_preparation_1997}, we can express the initial cavity field state in the Fock basis as $\ket{\psi}_{\mathrm{c}} = \sum_{n} c_{n} \ket{n}$. The mirror's initial state is introduced with the use of the Glauber-Sudarshan P representation~\cite{gerry_introductory_2023}:
\begin{equation}
 \ket{\phi}\bra{\phi}_{\mathrm{m}} = \int_{-\infty}^{+\infty} P_0(\beta) \ket{\beta}\bra{\beta} \mathrm{d}^2 \beta.   
\end{equation}
The solution to the Liouville–von Neumann equation can be expressed as
\begin{align}
	\hat{\rho}(t)=&\nonumber \sum_{n,m}  c_{n}c^{*}_{m}\mathrm{e}^{\mathrm{i} k^2 (n^2-m^2)(t-\sin t)}\ \ket{n}\bra{m}_{\mathrm{c}}\otimes I,
\end{align}
where
\begin{equation}
   I=\int_{-\infty}^{+\infty} P_0(\beta) \hat{D}_n(t) \ket{\beta \mathrm{e}^{-\mathrm{i}t}}\bra{\beta \mathrm{e}^{-\mathrm{i}t}} \hat{D}^\dagger_m(t) \mathrm{d}^2 \beta ,
\end{equation}
with
\begin{equation}
   \hat{D}_n(t) =\hat{D}[ kn(1-\mathrm{e}^{-\mathrm{i}t})] 
\end{equation}
being a form of the displacement operator. For an arbitrary initial mirror state $\ket{\phi}_{m}$ at time $t = 2 \pi M$, where $M$ is an integer multiple of the mirror's motion period $2 \pi$, the cavity field and mirror states can be separated, as they are not entangled at this moment.

To prepare $L$-component cat states with the use of the optomechanical setup described earlier, we begin by setting the initial state of the cavity field to a coherent state $\ket{\psi}_{\mathrm{c}} = \ket{\alpha_0}$. At separable moments, the state of the cavity field is given as
	\begin{equation}\label{eq:28}
	\ket{\psi(t =  2 \pi \times M)}_{\mathrm{c}} =  \mathrm{e}^{-\frac{|\alpha_0|^2}{2}} \sum_{n=0}^{\infty}  \frac{\alpha_0^n}{\sqrt{n !}} \mathrm{e}^{\mathrm{i} 2 M \pi k^2n^2}\ket{n}.
	\end{equation}
Note that the parameter $k$ is crucial throughout the whole temporal evolution. Rewriting Eq.\;(\ref{eq:28}) involves use of the following identities~\cite{agarwal_fractional_1998,robinett_quantum_2004}:
\begin{equation}\label{eq:frac}
\begin{aligned}
\text{for all } n \in \mathbb {N} , \quad &\mathrm{e}^{\mathrm{i} 2 \pi \frac{p}{q} n^2 } = \sum_{s =0}^{l-1} b_s \mathrm{e}^{ -\mathrm{i} 2 \pi \frac{s}{l} n},\\
& b_s := \frac{1}{l} \sum_{j =0}^{l-1} \mathrm{e}^{ \mathrm{i} 2 \pi\left( \frac{s}{l}j+ \frac{p}{q}j^2 \right) },
\end{aligned}
\end{equation}
where $p$ and $q$ are coprime integers, and $l = \nicefrac{q}{2}$ if $q$ is a multiple of $4$, otherwise $l = q$.

At time $t=2 \pi M $, when $M k^2\equiv  \frac{p}{q}~(\text{mod }1)$ is satisfied, the wave function of the cavity system splits into multiple coherent-state superpositions, 
\begin{equation}\label{eq:get_cat_states}
	\ket{\psi(t= 2 \pi M)}_{\mathrm{c}} = \sum_{s=0}^{l-1}  b_s \ket{\alpha_0 \mathrm{e}^{ -\mathrm{i} 2 \pi \frac{s}{l}} },
\end{equation}
with the modulus of the expansion coefficient
\begin{equation}\label{eq:norm}
\left\lvert b_s \right\rvert=
\begin{cases}
\sqrt{\frac{2}{q}}\delta_{s,\mathrm{odd}},\quad  q \equiv 2(\mathrm{mod }~4), \\
\sqrt{\frac{2}{q}},\quad  q \equiv 0(\mathrm{mod }~4),\\
\sqrt{\frac{1}{q}},\quad   q \text{ is odd}.
\end{cases}
\end{equation}
The initial coherent wave packet divides into $q$ wave packets if $q$ is odd, or $\nicefrac{q}{2}$ wave packets if $q$ is even, all with equivalent amplitude.

\begin{figure}
\includegraphics[width=0.51\textwidth]{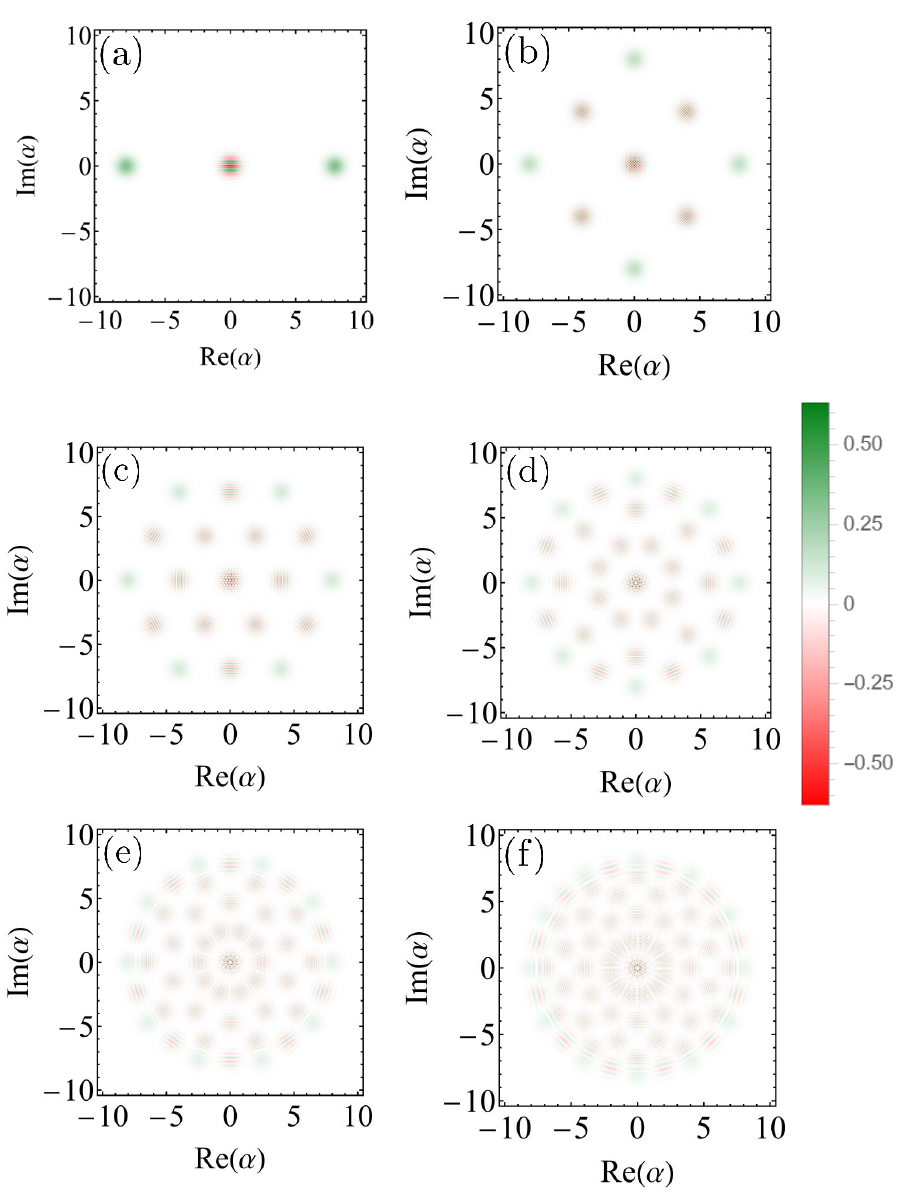}
\caption{Wigner distributions for optomechanical analogues, as described by Eqs.~(\ref{opto_2compo})– (\ref{opto_12compo}), showing phase-space representations for: (a) two-component, (b) four-component, (c) six-component, (d) eight-component, (e) ten-component, and (f) 12-component cat states.}
\label{fig:fig12}
\end{figure}

In this study, we keep the coupling constant $k$ fixed at $\nicefrac{1}{\sqrt{240}}$ throughout the evolution, while varying the parameter $M$. Different values of $M$ lead to different time periods, causing the wave packet to evolve into specific quantum states. We focus on $M$ values corresponding to the quantum state of interest, and examine the associated Wigner functions, observing their evolution as a function of $M$.

When $M=60$, Eq.~(\ref{eq:28}) clearly shows a superposition of two coherent states, demonstrating that the optomechanical setup can produce a two-component macroscopic cat state. This instantaneous quantum state takes the form
\begin{equation}\label{opto_2compo}
\ket{\psi(t=2\pi \times 60)}_{\mathrm{c}} = \frac{1}{\sqrt{2}}\Big( \ket{\alpha_0} +\text{e}^{\text{i}\pi}\ket{-\alpha_0}\Big).
\end{equation}
It is observed that a four-component cat state (compass state) arrives at $M=30$:
\begin{align}
\ket{\psi(t=2\pi\times30)}_{\mathrm{c}}= &\nonumber \frac{1}{2} \Big(\ket{\alpha_0} + \mathrm{e}^{-\mathrm{i}\frac{\pi}{4}} \ket{- \mathrm{i} \alpha_0 }  +\text{e}^{\text{i}\pi} \ket{- \alpha_0 }  \\&+ \mathrm{e}^{-\mathrm{i}\frac{\pi}{4}}\ket{\mathrm{i} \alpha_0 } \Big).
\end{align}
Similarly, 6-, 8-, 10-, and 12-component cat states correspond to \( M \) values of 20, 15, 12, and 10, respectively. 

An illustration of the six-component cat state analogue is shown as
\begin{align}
    &\nonumber\ket{\psi(t=2\pi\times20)}_{\mathrm{c}}=\\&\nonumber\frac{1}{\sqrt{6}} \Big(\ket{\alpha_0} +  \mathrm{e}^{-\mathrm{i}\frac{\pi}{6}} \ket{\alpha_0 \mathrm{e}^{-\mathrm{i}\frac{\pi}{3}}} +  \mathrm{e}^{-\mathrm{i}\frac{2\pi}{3}} \ket{\alpha_0 \mathrm{e}^{-\mathrm{i}\frac{2\pi}{3}}}
\\&+  \mathrm{e}^{\text{i}\frac{\pi}{2}} \ket{-\alpha_0}  + \mathrm{e}^{-\mathrm{i}\frac{2\pi}{3}}\ket{\alpha_0 \mathrm{e}^{\mathrm{i}\frac{2\pi}{3}}} + \mathrm{e}^{-\mathrm{i}\frac{\pi}{6}} \ket{\alpha_0 \mathrm{e}^{\mathrm{i}\frac{\pi}{3}}}\Big).
\end{align}
For the eight-component cat state, we have
\begin{align}
    &\nonumber\ket{\psi(t=2\pi\times 15)}_{\mathrm{c}} =\\&\nonumber\frac{1}{\sqrt{8}}\Big( \ket{\alpha_0} + \mathrm{e}^{-\mathrm{i}\frac{\pi}{8}} \ket{\alpha_0 \mathrm{e}^{-\mathrm{i}\frac{\pi}{4}}} + \mathrm{e}^{\text{i}\pi} \ket{- \mathrm{i}\alpha_0} + \mathrm{e}^{-\mathrm{i}\frac{9\pi}{8}} \ket{\alpha_0 \mathrm{e}^{-\mathrm{i}\frac{3\pi}{4}}} \\
&+ \ket{-\alpha_0} + \mathrm{e}^{-\mathrm{i}\frac{9\pi}{8}} \ket{\alpha_0 \mathrm{e}^{\mathrm{i}\frac{3\pi}{4}}} -\mathrm{e}^{\text{i}\pi}\ket{\mathrm{i}\alpha_0} + \mathrm{e}^{-\mathrm{i}\frac{\pi}{8}}\ket{\alpha_0 \mathrm{e}^{\mathrm{i}\frac{\pi}{4}}}\Big).
\end{align}
This configuration leads to the generation of analogues of 10- and 12-component cat states:
\begin{widetext}
\begin{align}
\ket{\psi(t=2\pi\times12)}_{\mathrm{c}} =
&\nonumber\frac{1}{\sqrt{10}}\Bigg[ \ket{\alpha_0} +  \mathrm{e}^{-\mathrm{i}\frac{\pi}{10}} \ket{\alpha_0 \mathrm{e}^{-\mathrm{i}\frac{\pi}{5}}} +   \mathrm{e}^{-\mathrm{i}\frac{2\pi}{5}} \ket{\alpha_0 \mathrm{e}^{-\mathrm{i}\frac{2\pi}{5}}}  + \mathrm{e}^{-\mathrm{i}\frac{9\pi}{10}} \ket{\alpha_0 \mathrm{e}^{-\mathrm{i}\frac{3\pi}{5}}}\\
& +\nonumber \mathrm{e}^{\mathrm{i}\frac{2\pi}{5}} \ket{\alpha_0 \mathrm{e}^{-\mathrm{i}\frac{4\pi}{5}}} + \mathrm{e}^{\text{i}\pi}\ket{-\alpha_0} + \mathrm{e}^{\mathrm{i}\frac{2\pi}{5}} \ket{\alpha_0 \mathrm{e}^{\mathrm{i}\frac{4\pi}{5}}} +  \mathrm{e}^{-\mathrm{i}\frac{9\pi}{10}} \ket{\alpha_0 \mathrm{e}^{\mathrm{i}\frac{3\pi}{5}}}\\
& +\mathrm{e}^{-\mathrm{i}\frac{2\pi}{5}} \ket{\alpha_0 \mathrm{e}^{\mathrm{i}\frac{2\pi}{5}}} + \mathrm{e}^{-\mathrm{i}\frac{\pi}{10}} \ket{\alpha_0 \mathrm{e}^{\mathrm{i}\frac{\pi}{5}}}\Bigg],
\end{align}
\end{widetext}
\begin{widetext}
\begin{align}\label{opto_12compo}
\ket{\psi(t=2\pi\times10)}_{\mathrm{c}} =
&\nonumber\frac{1}{\sqrt{12}} \Bigg[\ket{\alpha_0} +\mathrm{e}^{ -\mathrm{i} \frac{\pi}{12}} \ket{\alpha_0 \mathrm{e}^{ -\mathrm{i} \frac{\pi}{6}}} + \mathrm{e}^{ -\mathrm{i} \frac{\pi}{3}}\ket{\alpha_0 \mathrm{e}^{ -\mathrm{i} \frac{\pi}{3}}} + \mathrm{e}^{ -\mathrm{i} \frac{3\pi}{4}} \ket{-\mathrm{i} \alpha_0} \\
&\nonumber+ \mathrm{e}^{ \mathrm{i} \frac{2\pi}{3}} \ket{\alpha_0 \mathrm{e}^{ -\mathrm{i} \frac{2\pi}{3}}} + \mathrm{e}^{ -\mathrm{i} \frac{\pi}{12}} \ket{\alpha_0 \mathrm{e}^{ -\mathrm{i} \frac{5\pi}{6}}} +\mathrm{e}^{\text{i}\pi}  \ket{-\alpha_0} + \mathrm{e}^{ -\mathrm{i} \frac{\pi}{12}} \ket{\alpha_0 \mathrm{e}^{ \mathrm{i} \frac{5\pi}{6}}}\\
&  +\mathrm{e}^{ \mathrm{i} \frac{2\pi}{3}}\ket{\alpha_0 \mathrm{e}^{ \mathrm{i} \frac{2\pi}{3}}} + \mathrm{e}^{\mathrm{i} \frac{3\pi}{4}} \ket{\mathrm{i} \alpha_0} + \mathrm{e}^{ -\mathrm{i} \frac{\pi}{3}}\ket{\alpha_0 \mathrm{e}^{\mathrm{i} \frac{\pi}{3}}} + \mathrm{e}^{ -\mathrm{i} \frac{\pi}{12}} \ket{\alpha_0 \mathrm{e}^{\mathrm{i} \frac{\pi}{6}}}\Bigg].
\end{align}
\end{widetext}

The superpositions described in Eqs.~(\ref{opto_2compo})–(\ref{opto_12compo}) provide a clear and direct analogy to the superposition states discussed in Sec.~ \ref{sec:superpositions}. Each of the superpositions presented in these equations corresponds to a specific instance of the quantum states outlined earlier, illustrating how the mathematical form and underlying properties of these states are preserved when translated into the context of the optomechanical system. In Fig.~\ref{fig:fig12}, we show the Wigner distributions for each of the cases, highlighting their phase-space features and implying that these features are nearly similar to those discussed in Sec.~ \ref{sec:superpositions}. 
This indicates that the optomechanical system effectively generates near versions of the desired quantum states.

\begin{figure}
\centering
\includegraphics[width=0.5\textwidth]{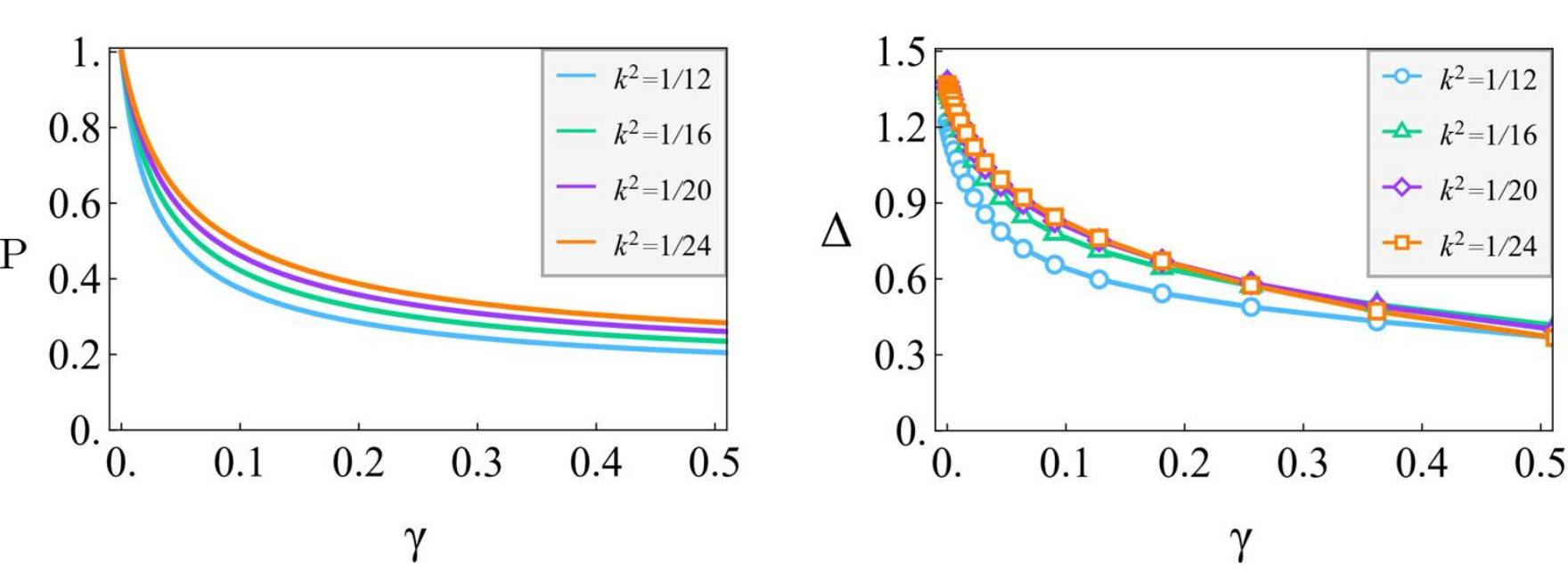}
\caption{Purity and Wigner negativity $\Delta$ of the cavity state $\hat{\rho}^{(\gamma)}_{\mathrm{c}}$ in the presence of mechanical dissipation. The coupling parameter $k = \nicefrac{1}{\sqrt{12}}$, $k = \nicefrac{1}{\sqrt{16}}$, $k = \nicefrac{1}{\sqrt{20}}$, and $k = \nicefrac{1}{\sqrt{24}}$ corresponds to the $6$-, $8$-, $10$-, and $12$-component cat states, respectively, and each observation is performed under unitary evolution.}
\label{fig:gamma}
\end{figure}

\begin{figure}
\centering
\includegraphics[width=0.5\textwidth]{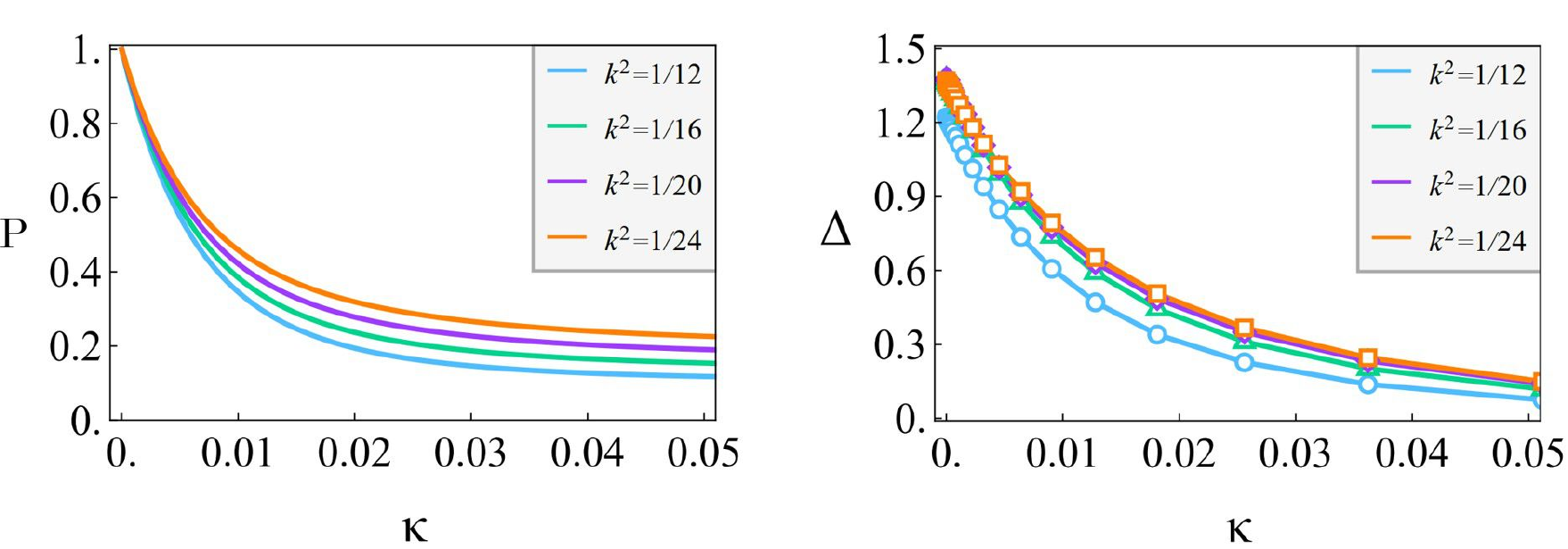}
\caption{Purity and Wigner negativity $\Delta$ of the cavity state $\hat{\rho}^{(\kappa)}_{\mathrm{c}}$ when there is cavity intensity decay in Eq.~(\ref{eq:master_eq2}).  The coupling parameter $k = \nicefrac{1}{\sqrt{12}}$, $k = \nicefrac{1}{\sqrt{16}}$, $k = \nicefrac{1}{\sqrt{20}}$, and $k = \nicefrac{1}{\sqrt{24}}$  corresponds to $6$-, $8$-, $10$-, and $12$-component cat states, respectively, under unitary evolution.}
\label{fig:kappa}
\end{figure}

\subsection{Significant decoherence}
Here we examine the decoherence effects in our optomechanical setup by using the standard master equations presented in Ref.~\cite{naseem_thermodynamic_2018}. These equations are derived on the basis of a series of approximations under the assumption of weak optomechanical coupling, which allows us to consider two thermal baths independently coupled to the optical cavity and the mechanical oscillator. Specifically, we solve the master equation for the time-dependent density matrix $\hat{\rho}^{(\gamma)}$ in the presence of mechanical dissipation at zero temperature and for $\hat{\rho}^{(\kappa)}$ in the case of cavity intensity decay at zero temperature.

We assume that the initial state at $t=0$ is a direct product of the cavity field state and the mirror state, both of which are coherent states of real-number amplitudes $\alpha_0$ and $\beta_0$; that is,

\begin{equation}
\hat{\rho}\left(t=0 \right) =  \ket{\alpha_0}\bra{\alpha_0}_{\mathrm{c}} \otimes \ket{\beta_0}\bra{\beta_0}_{\mathrm{m}}.
\end{equation}
Consider the situation when photon leakage in the cavity is minimal and the dissipation of the mirror motion is large. The corresponding master equation can be expressed as
\begin{equation}\label{eq:master_eq1}
	\frac{\mathrm{d} \hat{\rho}^{(\gamma)}(t)}{\mathrm{d} t}=- \mathrm{i}[\hat{H},  \hat{\rho}^{(\gamma)}]+\frac{\gamma}{2}\left[2 \hat{b}  \hat{\rho}^{(\gamma)} \hat{b} ^{\dagger}-\hat{b}^{\dagger}\hat{b}  \hat{\rho}^{(\gamma)}- \hat{\rho}^{(\gamma)} \hat{b}^{\dagger} \hat{b}\right].
\end{equation}
Here the mechanical dissipation rate $\gamma$ accounts for the phonon loss of the mechanical oscillator.  In addition to the well-known viscous damping, another significant source of dissipation is the clamping losses arising from the structural design of mirror devices~\cite{RevModPhys.86.1391}, which may dominate other loss mechanisms.

The time-dependent density matrix $\hat{\rho}^{(\gamma)}$ is obtained by our following the techniques presented in Ref.~\cite{bose_preparation_1997}, which specifically are based on the Trotter decomposition of superoperators. We rewrite the master equation in a vector notation~\cite{campaioli_quantum_2024}, $\frac{\mathrm{d}}{\mathrm{d} t }\ket{\hat{\rho}^{(\gamma)}} =\Tilde{\mathcal{L} }\ket{\hat{\rho}^{(\gamma)}} $, where $\ket{\hat{\rho}^{(\gamma)}} $ is the vectorized form of the density matrix $\hat{\rho}^{(\gamma)}$ in Fock-Liouville space and $\Tilde{\mathcal{L} }$ is the superoperator, which can be expressed as
\begin{equation}
\Tilde{\mathcal{L} } = -\mathrm{i}(\hat{I} \otimes \hat{H}-\hat{H}^{\mathrm{T}} \otimes \hat{I})+ [\hat{b}^* \otimes \hat{b}-\frac{1}{2}(\hat{I} \otimes \hat{b}^{\dagger} \hat{b}+\hat{b}^{\mathrm{T}} \hat{b}^* \otimes \hat{I})],
\end{equation}
where the superoperator $\Tilde{\mathcal{L} }$ can be divided into the sum of two parts $\Tilde{\mathcal{L} } = \Tilde{\mathcal{L}}_A + \Tilde{\mathcal{L} }_B $, where $\Tilde{\mathcal{L}}_A$ represents unitary evolution and $\Tilde{\mathcal{L}}_B$ represents nonunitary evolution.  The vectorized form of the resulting density matrix is given by $\ket{\hat{\rho}^{(\gamma)}(t)} = \mathrm{e}^{( \Tilde{\mathcal{L}}_A + \Tilde{\mathcal{L} }_B)t}\ket{\hat{\rho}^{(\gamma)}(0)}$, and using the Trotter decomposition, we rewrite $\mathrm{e}^{( \Tilde{\mathcal{L}}_A + \Tilde{\mathcal{L} }_B)t} = \lim_{N\to + \infty} (\mathrm{e}^{ \Tilde{\mathcal{L}}_A \frac{t}{N}}\mathrm{e}^{\Tilde{\mathcal{L} }_B\frac{t}{N}})^N $, allowing us to obtain the solution of the master equation using a way in which unitary and nonunitary evolution proceeds alternately in infinitesimal time. The time-dependent density matrix $\hat{\rho}^{(\gamma)}$ is obtained as
\begin{equation}
\begin{aligned}
	\hat{\rho}^{(\gamma)}  =&  \mathrm{e}^{-|\alpha_{0}|^2}\sum_{n,m=0}^{+\infty} \Bigg[ \frac{\alpha_{0}^{n+m}}{\sqrt{n ! m!}} \mathrm{e}^{-D_{nm}(\gamma,t)}\mathrm{e}^{\mathrm{i}K_{nm}(\gamma,t)}   \\
	&\times \ket{n}\bra{m}_{\mathrm{c}} \otimes   \ket{\phi_n(\gamma,t)}\bra{\phi_m(\gamma,t)}_{\mathrm{m}}\Bigg] ,
    \end{aligned}
\end{equation}	
where $\ket{\phi_n(\gamma,t)}_{\mathrm{m}}$ is the coherent state of the mirror with amplitude $\phi_n(\gamma,t)$ given by
\begin{equation}
\begin{aligned}
&\phi_n(\gamma,t)\\
=&\mathrm{e}^{-\frac{\gamma t}{2}}\left[ \beta _0 \cos t+\frac{2 k n}{\gamma^2 +4} \left(2 e^{\frac{\gamma  t}{2}}-\gamma  \sin t -2 \cos t\right) \right] \\
&+\mathrm{i}\mathrm{e}^{-\frac{\gamma t}{2}}\left[-\beta _0  \sin t +\frac{2 k n}{\gamma^2 +4} \left(\gamma  e^{\frac{\gamma  t}{2}}-\gamma  \cos t+2 \sin t\right) \right].
\end{aligned}
\end{equation}
We use the function $D_{nm}(\gamma, t)$, which reflects the quantum coherence decay between Fock states; that is,
\begin{align}
	D_{nm} (\gamma,t) &\nonumber=\frac{\gamma}{2}\int_{0}^{t} \lvert\phi_n(\gamma, \tau ) - \phi_m(\gamma, \tau ) \rvert^2 \mathrm{d} \tau  \\&\nonumber=\frac{ 2 k^2 (m-n)^2 }{\gamma ^2+4}\Bigg[ \gamma t+ 1-\mathrm{e}^{-\gamma  t }-\frac{4 \gamma \mathrm{e}^{-\frac{\gamma  t}{2}  }}{\gamma ^2+4}\\&\times\Big(\gamma  \mathrm{e}^{\frac{\gamma  t }{2}}-\gamma  \cos t +2 \sin t \Big) \Bigg].
\end{align}
The real-valued function $K_{nm}(\gamma, t)$ denotes the relative phases between Fock states, and it can be written as
\begin{widetext}
\begin{align}
K_{nm}  (\gamma,t) = &\nonumber k \int_{0}^{t}  n \mathrm{Re}\left[\phi_n(\gamma, \tau ) \right] - m \mathrm{Re}\left[\phi_m(\gamma, \tau ) \right] \mathrm{d} \tau  \\
 = -&\nonumber\frac{2 \beta _0 k (m - n) \mathrm{e}^{-\frac{\gamma  t}{2}}}{\gamma^2 + 4}  \left( \gamma  \mathrm{e}^{\frac{\gamma  t}{2}}-\gamma  \cos t+2 \sin t\right)+ \frac{4 k^2 ( m^2-n^2) }{(\gamma ^2+4)^2} \Big[ 4 \gamma -\left(\gamma ^2+4\right) t \\
\quad - &\mathrm{e}^{-\frac{\gamma  t}{2} } \left(\gamma ^2 \sin t-4 \sin t+4 \gamma  \cos t\right)\Big].
\end{align}
\end{widetext}
We now solely consider the effects associated with the leakage of the cavity, while the dissipation of the mirror is assumed to be negligible. In this case, we have
\begin{equation}\label{eq:master_eq2}
	\frac{\mathrm{d} \hat{\rho}^{(\kappa)}}{\mathrm{d} t}=- \mathrm{i}[\hat{H},  \hat{\rho}^{(\kappa)}]+\frac{\kappa}{2}\left[2 \hat{a}  \hat{\rho}^{(\kappa)} \hat{a} ^{\dagger}-\hat{a}^{\dagger}\hat{a}  \hat{\rho}^{(\kappa)}- \hat{\rho}^{(\kappa)} \hat{a}^{\dagger} \hat{a}\right].
\end{equation}
Here the cavity intensity decay rate $\kappa$ accounts for the photon loss. It has two sources: one related to input or output coupling, and the other being the loss inside the cavity, such as that caused by the flatness and roughness of the mirror surface\cite{straniero_realistic_2015}.

Using the same method as used in Ref.~\cite{skvarcek_phase_1999} and then following the series solution of the master equation, we get time-dependent density matrix $\hat{\rho}^{(\kappa)}$ as
\begin{align}
\hat{\rho}^{(\kappa)}  =&\nonumber \mathrm{e}^{-|\alpha_{0}|^2}  \sum_{n,m=0}^{+\infty}\Bigg[\frac{\alpha_{0}^{n+m}}{\sqrt{n!m!}}   \rho_{nm} (\kappa,t)
 \\ &\times \ket{n}\bra{m}_{\mathrm{c}} \otimes   \ket{\phi_n(0,t)}\bra{\phi_m(0,t)}_{\mathrm{m}}\Bigg],
\end{align}
with
\begin{equation}
	\begin{aligned}
		\rho_{nm} (\kappa,t) =& \mathrm{e}^{-\frac{\kappa}{2}(n+m)t}\mathrm{e}^{\mathrm{i}k^2(n^2-m^2)( t - \sin t )} \\ &\times
        \exp\left(\kappa \alpha_{0}^2 \int_{0}^{t}  \mathrm{e}^{-\kappa \tau +2 \mathrm{i}k^2(n-m)(\tau-\sin \tau)} \mathrm{d}\tau \right).
	\end{aligned}
\end{equation}
This shows that in this case the states of the cavity and the mirror are mixed when $t \ne 0$. To obtain a complete description of the cavity system, we calculate the partial trace of the optomechanical system to get the density matrix $\hat{\rho}_{\mathrm{c}}^{(i)}$; that is,
\begin{equation}
	\hat{\rho}_{\mathrm{c}}^{(i)}
    =\mathrm{Tr}_{\mathrm{m}}	\hat{\rho}^{(i)},\quad i= \gamma,\kappa.
\end{equation}
Mathematically, the purity of the state $\hat{\rho}_{\mathrm{c}}^{(i)}$ can be measured by using
\begin{equation}\label{eq:pure}
\text{P}:=\mathrm{Tr}\Big( \hat{\rho}_{\mathrm{c}}^{(i)} \Big)^2,
\end{equation}
which alternatively also quantifies the amount of mixedness in a state.

To analyze the decoherence effects, we measure the loss of Wigner negativity and the purity associated with our higher-order superposition. The negativity of the Wigner function for the density matrix of the cavity system $\hat{\rho}^{(i)}_{\mathrm{c}}$ is defined as~\cite{kenfack_negativity_2004}
	\begin{equation}\label{eq:wig_neg}
		\Delta:= \frac{1}{2}\int_{-\infty}^{+\infty}| W_{\hat{\rho}^{(i)}_{\mathrm{c}}}(\alpha) |\mathrm{d}^2\alpha - \frac{1}{2}.
	\end{equation}
Sub-Planckian features observed in our states appear to be crucial for the enhanced susceptibility to perturbation possessed by our proposed quantum states. These phase-space features are prone to environmental decoherence, which can be quantified by the observation of a decrease in the Wigner negativity to decoherence.

\begin{figure}
\centering
\includegraphics[width=0.5\textwidth]{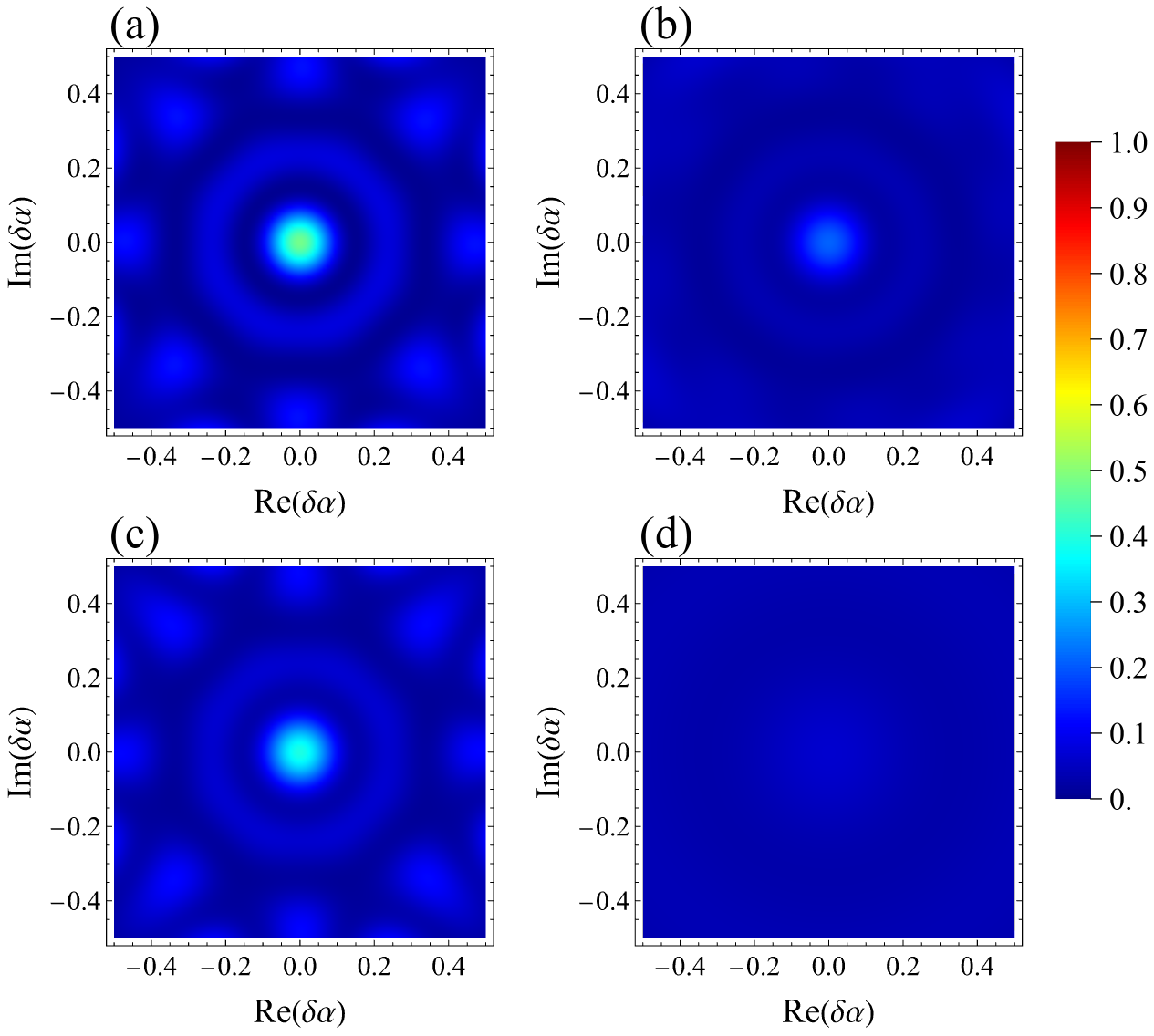}
\caption{The overlap function $O(\delta x, \delta p)$ of the mixed variants, denoted as $\hat{\rho}^{(\gamma)}$ and $\hat{\rho}^{(\kappa)}$, evolves from an eight-component cat state: (a) $\gamma = 0.02$, $\kappa = 0$; (b) $\gamma = 0.2$, $\kappa = 0$; (c) $\gamma = 0$, $\kappa = 10^{-4}$; (d) $\gamma = 0$, $\kappa = 10^{-3}$. In all situations, the initial state is specified as $\alpha_0 = 8$, $\beta_0 =0$, the coupling parameter $k=\nicefrac{1}{\sqrt{240}}$, and the evolution time $t=2\pi \times 15$.}

\label{fig:mixed_overlaps}
\end{figure}

\begin{figure}
\centering
\includegraphics[width=0.47\textwidth]{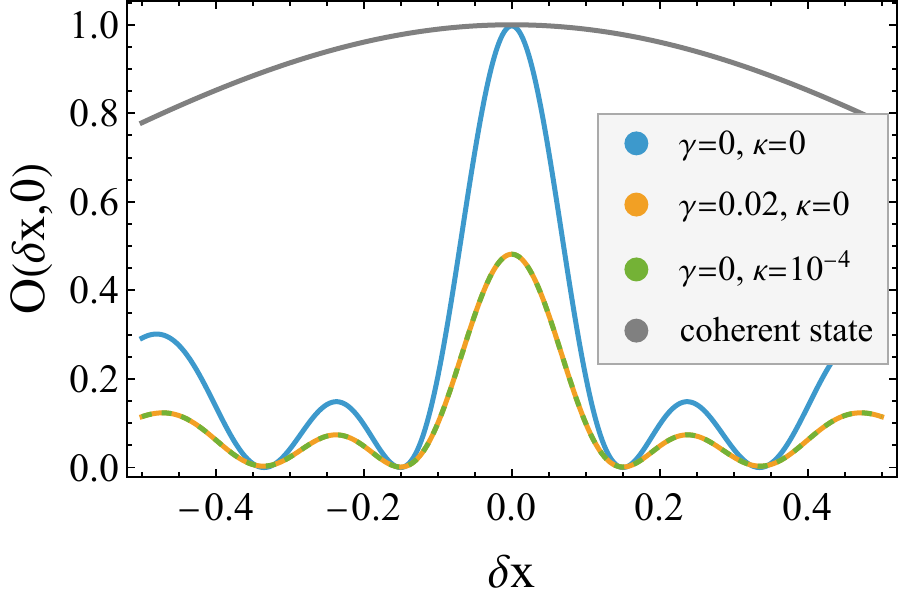}
\caption{Overlap function for different sets of states, $O(\delta x, \delta p)$, plotted along the $x$ direction in phase space (with $\delta p = 0$). The gray curve represents the coherent state, while the other curves correspond to eight-component cat states under various parameter selections.
}
\label{fig:slice_overlaps}
\end{figure}

To demonstrate the effect of decoherence, we quantify purity [Eq.~(\ref{eq:pure})] and Wigner negativity [Eq.~(\ref{eq:wig_neg})]. Note that these numerical representations are rather hard in our instances because of the increased number of coherent states in the superpositions. We set $\alpha_{0}=4$ and $\beta_{0}=0$ for a fixed evolution time $t = 2\pi$ and varied values of the coupling constant $k$. Under unitary evolution, the coupling constant $k$ plays a crucial role in determining the number of components in the superpositions. Specifically, $ k = \nicefrac{1}{\sqrt{12}}$
results in a six-component cat state. Similarly, $ k = \nicefrac{1}{\sqrt{16}}$ produces an 
eight-component cat state, $k = \nicefrac{1}{\sqrt{20}}$ yields a ten-component cat state, and $k = \nicefrac{1}{\sqrt{24}}$ gives rise to a 12-component cat state. 
Figures~\ref{fig:gamma} and \ref{fig:kappa} illustrate the purity and Wigner negativity of the corresponding states for varied $\gamma$ and $\kappa$, respectively. As $\gamma$ or $\kappa$ becomes substantial, the related states lose their quantum natures, reflected by the loss of purity and the associated Wigner negativity. The observed decay in the corresponding curves reflects a loss of purity and a reduction in the negative amplitudes of the Wigner function, indicating a gradual degradation of quantum coherence. As a result, key advantages such as enhanced sensitivity to displacements diminish, making the states less effective compared with their fully coherent quantum counterparts.

We observe that the enhancement in sensitivity exhibited by our quantum states, as quantified by their overlap functions, disappears in the presence of decoherence. This indicates a substantial loss of sub-Planckian sensitivity, with decoherence having a significant impact. This suggests that the retention of sub-Planckian sensitivity is compromised when decoherence effects become pronounced. This behavior is illustrated in Figs.~\ref{fig:mixed_overlaps} and \ref{fig:slice_overlaps}, where we set $\alpha_{0} = 8$ and $\beta_{0} = 0$ for a fixed evolution time $t = 2\pi \times 15$, while varying $\gamma$ and $\kappa$. We present the overlap functions under both mild and strong decoherence, using eight-component cat states under unitary evolution as a reference. The results show that the central peak in the oscillatory pattern, which signifies sub-Planckian sensitivity, progressively diminishes, indicating a loss of the enhanced isotropic sensitivity characteristic of the quantum state before decoherence. Importantly, we note that decoherence does not alter the overall shape of the structures, whether isotropic or anisotropic. The original structural patterns are retained but become increasingly blurred as a result of decoherence effects. Note that setting $\gamma = \kappa = 0$ results in the overlap of the eight-component cat state, indicating that our optomechanical analogues exhibit similar enhancement in sensitivity as their counterparts discussed in Sec.~\ref{sec:superpositions}.

\section{Conclusion AND REMARKS}

In conclusion, we explore superpositions of $L$ coherent states and investigate their phase spaces using the Wigner function formalism, demonstrating that for $L\ge6$ and even, these superpositions exhibit isotropic sub-Planckian structures in phase space. This isotropy extends to the sensitivity to phase-space displacements. Previously identified superpositions of compass states~\cite{shukla_superposing_2023} are a particular case of our generalized illustrations, showing that isotropic sub-Planckian structures can also be achieved through superpositions of simple macroscopic cat states. We proposed an optomechanical framework that generates analogues of the quantum states we suggested, thereby demonstrating the potential for the physical realization of our generic superpositions. We also discussed the decoherence effects associated with the optomechanical setup and noted that the nonclassical characteristics of our cases decay rapidly with larger $\gamma$ and $\kappa$.

Real optomechanical systems typically have significantly higher photon frequencies, ranging from $10^3$ to $10^6$ times that of mechanical oscillators~\cite{RevModPhys.86.1391}. This difference in energy scales raises two key concerns: First, the photon and the mechanical oscillator have very different frequencies, showing that their dynamics are influenced by distinct energy scales. The behavior of the optical cavity is influenced by photons, which have significantly higher energy, whereas optomechanical coupling has a greater effect on mechanical oscillators, which have less energy. 
Because the time frames and energy transfers differ by orders of magnitude, it is difficult to effectively reproduce the interactions between the two systems. Second, light from an outside source is frequently pumped into the optical cavity to create its modes. The optomechanical coupling constant $k$ is usually much less than the relaxation time of the optical cavity, which is related to the cavity transmission rate. This shows that the relaxation dynamics of the optical cavity affect its interaction with the mechanical oscillator, making it essential to consider these dynamics when one is evaluating the behavior of the system.

In a potential experimental realization, the overlap function can be coupled to the system with a two-level atom and read out via the excitation probability of the atom \cite{toscano_sub-planck_2006}. In our analysis, the optomechanical single-photon coupling strength $k$ does not necessarily require strong coupling (referring to the condition where the coupling parameter $k$ is greater than both the mechanical dissipation rate $\gamma$ and the cavity intensity decay rate $\kappa$). However, accurately setting of the coupling parameter $k$ is challenging, and the phenomenon may be disrupted by high-order optomechanical interactions not  included in our Hamiltonian [Eq.~(\ref{hami})]. Alternatively, multicomponent cat states have been created by trapped-ion methods~\cite{Li-Xiang,Zheng01031999,PhysRevA.58.761}, which possibly can also be adjusted particularly to generate the cat states we indicated. For example, in the work reported in Ref.~\cite{Zheng01031999}, a motional superposition of equidistant coherent states arranged on a circle, resulting in multicomponent cat states similar to ours, was achieved through the motion of a trapped ion with the use of a single conditional measurement. Additionally, such setups have been experimentally realized~\cite{johnson2017ultrafast} to generate superpositions of coherent states and perhaps can be further extended to the quantum states we presented.

Importantly, the existence of the sub-Planckian features has important ramifications for the sensitivity of quantum systems to external perturbations. Such sensitivity suggests promising applications in precision measurement and detection, most notably in the detection of weak-forces~\cite{toscano_sub-planck_2006}. In addition to weak force detection, potential applications may include magnetometry with enhanced sensitivity in the measurement of weak magnetic fields, possibly contributing to advancements in medical imaging and fundamental physics experiments, inertial sensing with high-precision accelerometers and gyroscopes based on the principles of sub-Planckian interference, possibly dramatically improving navigation systems, and gravitational wave detection. As the experimental techniques evolve, the integration of sub-Planckian sensitive systems into practical devices could lead to sensors with sensitivity that surpasses classical limits by orders of magnitude.
\section*{ACKNOWLEDGMENT}
This work was supported by the National Natural Science Foundation of China under Grant No.~12174346. N.A. acknowledges Xiaosen Yang at Jiangsu University for insightful comments on the work.
\section*{DATA AVAILABILITY}
The data that support the findings of this article are not
publicly available. The data are available from the authors
upon reasonable request.

\onecolumngrid
\appendix
\section{ARBITRARY SUPERPOSITION}

\begin{figure*}
\centering
\includegraphics[width=0.8\textwidth]{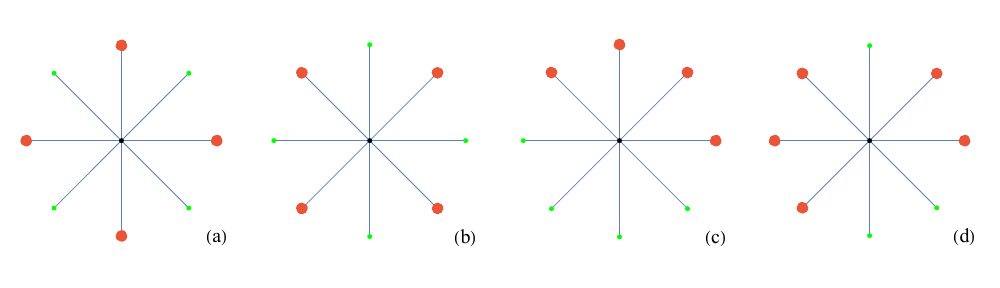}
\caption{Eight-component cat state with  uniform (nonuniform) phases (weights), where larger red dots and smaller tiny green dots represent coherent states with larger and smaller weights, respectively. Black dots represent the central phase-space component. (a)~$A_{0}=A_{2}=A_{4}=A_{6}=\nicefrac{2}{\sqrt{12}}$ and $A_{1}=A_{3}=A_{5}=A_{7}=\nicefrac{1}{\sqrt{12}}$. (b)~$A_{0}=A_{2}=A_{4}=A_{6}=\nicefrac{1}{\sqrt{12}}$ and  $A_{1}=A_{3}=A_{5}=A_{7}=\nicefrac{2}{\sqrt{12}}$.  (c)~$A_{0}=A_{1}=A_{2}=A_{3} = \nicefrac{2}{\sqrt{12}}$ and $A_{4}=A_{5}=A_{6}=A_{7} = \nicefrac{1}{\sqrt{12}}$. (d)~$A_{0}=A_{1}=A_{3}=A_{4}=A_{5} = \nicefrac{2}{\sqrt{13}}$ and $A_{2}=A_{6}=A_{7}= \nicefrac{1}{\sqrt{13}}$. In all cases, $\beta=8$ and $\omega_j = \nicefrac{2\pi j}{8}$.}
\label{fig:fig17}
\end{figure*}

\begin{figure*}
\centering
\includegraphics[width=0.8\textwidth]{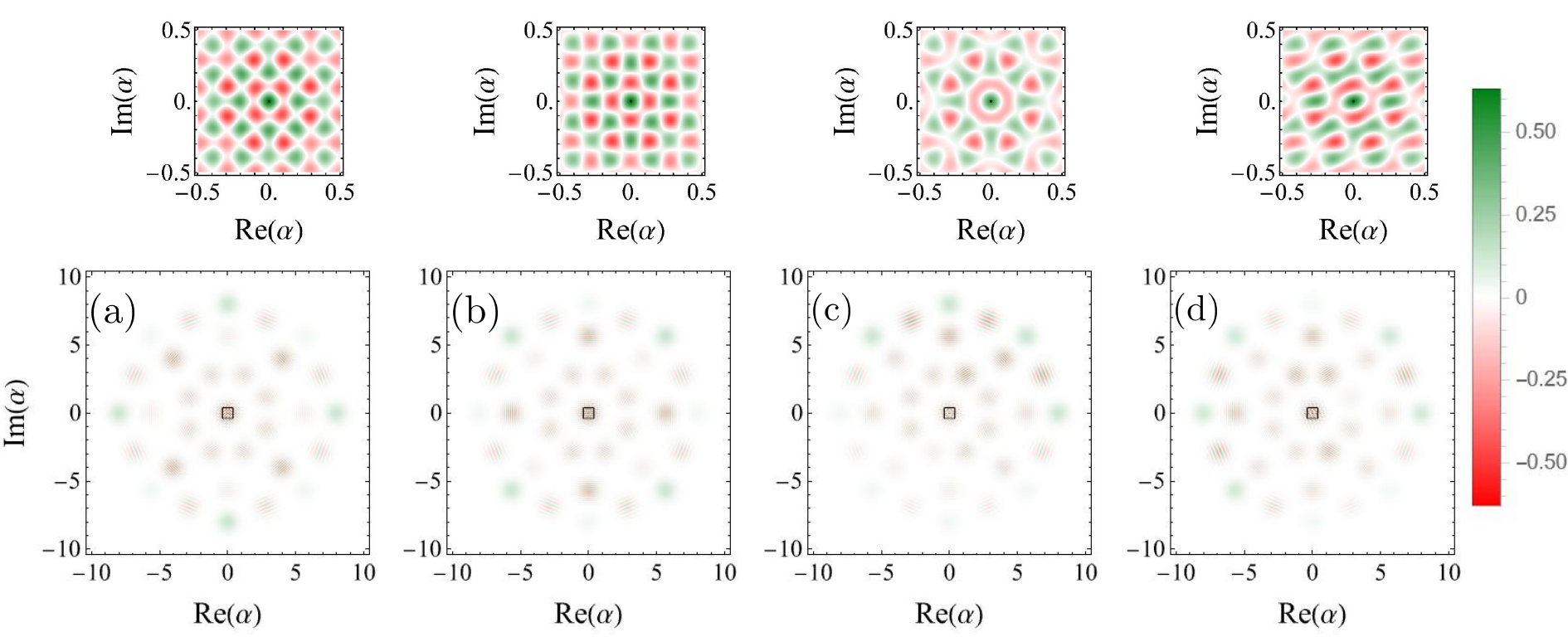}
\caption{Wigner distributions of eight-component
cat states with uneven weights for the cases presented in Fig.~\ref{fig:fig17} for (a) Fig.~\ref{fig:fig17}(a), (b) Fig.~\ref{fig:fig17}(b), (c) Fig.~\ref{fig:fig17}(c), and (d) Fig.~\ref{fig:fig17}(d). The insets illustrate a close-up of the central phase-space region.}
\label{fig:fig18}
\end{figure*}

\begin{figure*}
\centering
\includegraphics[width=0.8\textwidth]{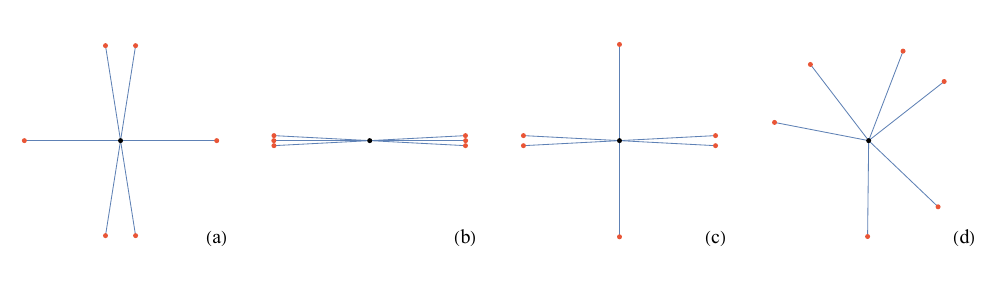}
\caption{Six-component cat state with the even weights and uneven phases for (a)~$\omega_0 = 0$, $\omega_1 = \nicefrac{9\pi}{20}$, $\omega_2 = \nicefrac{11\pi}{20}$, $\omega_3 = \pi$, $\omega_4 = -\nicefrac{11\pi}{20}$, and $\omega_5 = -\nicefrac{9\pi}{20}$, (b)~$\omega_0 = 0$, $\omega_1 = \nicefrac{\pi}{60}$, $\omega_2 = \nicefrac{59\pi}{60}$, $\omega_3 = \pi$, $\omega_4 = -\nicefrac{59\pi}{60}$, and $\omega_5 = -\nicefrac{\pi}{60}$, (c)~$\omega_0 = \nicefrac{\pi}{60}$, $\omega_1 = \nicefrac{\pi}{2}$, $\omega_2 = \nicefrac{59\pi}{60}$, $\omega_3 =-\nicefrac{59\pi}{60}$, $\omega_4 = -\nicefrac{\pi}{2}$, and $\omega_5 = -\nicefrac{\pi}{60}$, and (d)~six random phases for the six $\omega_j$ values of $0.664425$, $1.20411$, $2.22358$, $2.95041$, $4.70109$, and $5.52159$.}
\label{fig:fig19}
\end{figure*}

\begin{figure*}
\centering
\includegraphics[width=0.8\textwidth]{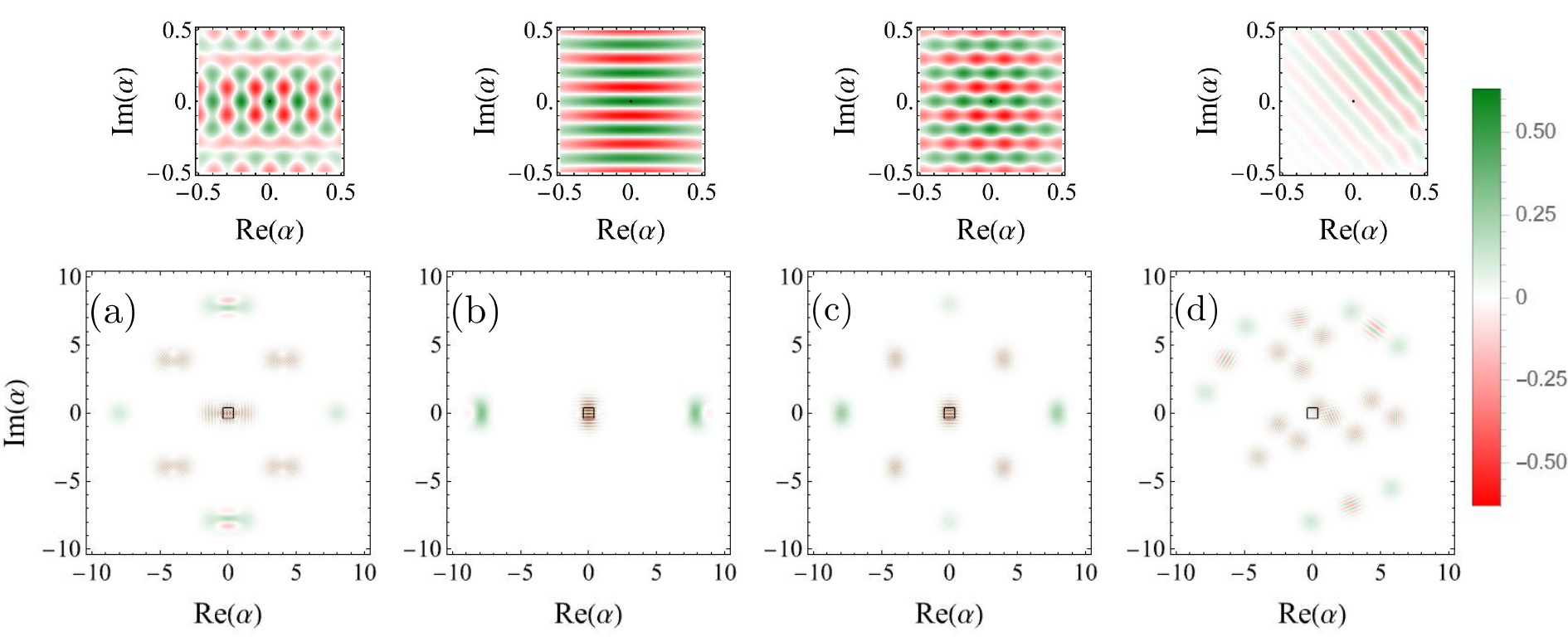}
\caption{Wigner distributions for six-component
cat states with uneven phases for the situations presented in (a) Fig.~\ref{fig:fig19}(a), (b) Fig.~\ref{fig:fig19}(b), (c) Fig.~\ref{fig:fig19}(c), and (d) Fig.~\ref{fig:fig19}(d), where the insets represent a close-up of the central phase-space region.}
\label{fig:fig20}
\end{figure*}

In the main text, we explored the superposition of coherent states with equal amplitudes, uniformly distributed in phase space, with these arrangements providing a circular pattern (for $L\geq4$) around the origin as depicted in Fig.~\ref{fig:fig1}. These illustrated cases represent the ideal scenarios for developing isotropic sub-Planckian features, which are directly linked to the isotropic enhancement in sensitivity. Now let us consider a more general form of such superpositions, where the component states have unequal spacing and weights:
\begin{equation}\label{eq:gener_super}
\ket{G_L}\propto\sum_{j=0}^{L-1} A_j\ket{\mathrm{e}^{\mathrm{i} \omega_j}\beta },
\end{equation}
with relative weights $A_{j}$  and phases $\omega_{j}$. To keep our discussion concise, we focus on a few superpositions of interest, constructed with different weights and/or phases, in order to analyze the effects on the phase-space features resulting from these deviations from the main text. This approach will also provide insights into other related superpositions.

First, we consider the case where the coherent states in the superposition (\ref{eq:gener_super}) have equal phases but unequal weights. We investigate this scenario by examining an eight-component cat state with the arrangement of coherent states shown in Fig.~\ref{fig:fig17}. In Fig.~\ref{fig:fig17}, the red and green dots represent the corresponding coherent states, with the green dots indicating the states with relatively smaller weights. First, consider the case shown in Fig.~\ref{fig:fig17}(a) and its corresponding Wigner distribution represented in Fig.~\ref{fig:fig18}(a). In this case, the coherent states located along the diagonals in phase space are weaker than the other components. As is clearly evident in the inset in Fig.~\ref{fig:fig17}(a), this pattern of the coherent states results in the central feature losing its isotropy, and it closely resembles the central feature of the four-component cat state. This highlights the significance of the diagonal components in forming the isotropic sub-Planckian structure. Figure~\ref{fig:fig17}(b) and the corresponding Wigner distributions in Fig.~\ref{fig:fig18}(b) represent the case in which the components of the coherent states in the superposition located along the horizontal and vertical axes in the phase space are now weaker than their diagonal components, simply providing the $\pi/4$-rotated case of Fig.~\ref{fig:fig17}(a). It is also apparent that anisotropic sub-Planckian features exist in this case. Now consider the case shown in Fig.~\ref{fig:fig17}(c), for which the corresponding Wigner distribution is shown in Fig.~\ref{fig:fig18}(c). In this case, the components of opposite phases for each pair of coherent states are weakening. Notably, this configuration maintains the isotropy of the sub-Planckian feature. In Fig.~\ref{fig:fig17}(d) and \ref{fig:fig18}(d), the previously uniform distribution of the coherent states is disrupted, leading to the emergence of a highly anisotropic region around the center of the phase space.

We now examine situations where the phases of the coherent states are random and do not follow the uniform pattern outlined in the main text. We find that deviations from the ideal phase choices, which were initially intended to create isotropic regions in phase space, cause significant disruption to these regions. This is clearly shown in Figs.~\ref{fig:fig20}(a)–\ref{fig:fig20}(d). This suggests that in order to create the isotropic sub-Planckian feature, it is essential to construct the superposition by distributing the components equally across the circular region in phase space. The choice of weights plays a crucial role, as the presence of components at opposite phases, even if weaker, ensures the formation of the isotropic features. It is important to note that the sensitivity of states is directly related to the structure of their phase-space features. Specifically, a quantum state with finer phase-space features exhibits a corresponding refinement in sensitivity, as described in Eq.~(\ref{eq:overlap}), and this can be easily understood in the context of each of the cases illustrated.
\twocolumngrid
\bibliography{ref}
\end{document}